\documentclass[superscriptaddress,amsmath,amssymb,longbibliography,aps]{revtex4-2}

\usepackage{graphicx}
\usepackage{overpic}
\usepackage[table]{xcolor}
\usepackage{dcolumn}% Align table columns on decimal point
\usepackage{bm}
\usepackage{hyperref}
\usepackage{natbib}
\usepackage{multirow}
\usepackage{siunitx}
\usepackage{tikz}
\usepackage{MnSymbol}
\usepackage{gensymb}
\usepackage{comment}
\usepackage{bbding}

\definecolor{darkred}{rgb}{0.4,0,0}
\definecolor{darkgreen}{rgb}{0,0.4,0}
\definecolor{darkblue}{rgb}{0,0,0.4}
\definecolor{linegreen}{rgb}{0,0.8,0}

\DeclareMathOperator{\ku}{Ku}

\newcommand{\tauK}{\tau_{\eta}}
\newcommand{\etaK}{\eta}
\newcommand\ve[1]{\boldsymbol{#1}}
\newcommand{\ma}[1]{\ensuremath{\mathbb{#1}}}

\newcommand{\omegasp}{\omega^{({\rm s})}_p}
\newcommand{\omegasq}{\omega^{({\rm s})}_q}
\newcommand{\omegas}{\omega^{({\rm s})}}

\newcommand{\vs}{v^{({\rm s})}}

\newcommand{\vsmax}{v^{({\rm s})}_{\rm max}}
\newcommand{\omegasmax}{\omega^{({\rm s})}_{\rm max}}
\newcommand{\vsopt}{v^{({\rm s})}_{\rm opt}}
\newcommand{\omegasopt}{\omega^{({\rm s})}_{\rm opt}}
\newcommand{\Tp}{T_{\rm p}}
\newcommand{\Tu}{T_{\rm u}}

\newcommand{\tr}{\ensuremath{\mbox{tr}}}
\newcommand{\trSS}{{{\mathcal S}^2}}
\newcommand{\trSSt}{{{\mathcal S}(t)^2}}

\newcommand{\trSSarg}[1]{{{\mathcal S}(#1)^2}}
\newcommand{\nhat}{{\hat{\ve n}}}
\newcommand{\phat}{{\hat{\ve p}}}
\newcommand{\qhat}{{\hat{\ve q}}}
\newcommand{\tauj}{\tau_{{\rm j}}}

\newcommand{\PhiK}{\Phi_\eta}
\newcommand{\ellf}{\ell_{{\rm f}}}
\newcommand{\tauf}{\tau_{{\rm f}}}
\newcommand{\uf}{u_{{\rm f}}}

\newcommand{\cross}{\times}

\begin{document}

\title{Short term vs. long term: optimization of microswimmer navigation on different time horizons}

\author{N.Mousavi}
\affiliation{Department of Physics, University of Gothenburg, SE-41296 Gothenburg, Sweden}
\author{J. Qiu}
\affiliation{Department of Physics, University of Gothenburg, SE-41296 Gothenburg, Sweden}
\affiliation{AML, Department of Engineering Mechanics, Tsinghua University, 100084 Beijing, China}
\author{L. Zhao}
\affiliation{AML, Department of Engineering Mechanics, Tsinghua University, 100084 Beijing, China}
\author{B. Mehlig}
\affiliation{Department of Physics, University of Gothenburg, SE-41296 Gothenburg, Sweden}
\author{K. Gustavsson}
\email{kristian.gustafsson@physics.gu.se}
\affiliation{Department of Physics, University of Gothenburg, SE-41296 Gothenburg, Sweden}

\begin{abstract}
We use reinforcement learning to find strategies that allow microswimmers in turbulence to avoid regions of large strain. This question is motivated by the hypothesis that swimming microorganisms tend to avoid such regions to minimise the risk of predation. We ask which local cues a microswimmer must measure to efficiently avoid such straining regions. We find that it can succeed without directional information, merely by measuring the magnitude of the local strain. However, the swimmer avoids straining regions more efficiently if it can measure the sign of local strain gradients. We compare our results with those of an earlier study [Mousavi {\em et al.} Phys. Rev. Res. {\bf 6}, L022034 (2024)] where a short-time expansion was used to find optimal strategies. We find that the short-time strategies work well in some cases but not in others. We derive a new theory that explains when the time-horizon matters for our optimisation problem, and when it does not. We find the strategy with best performance when the time-horizon coincides with the correlation time of the turbulent fluctuations. We also explain how the update frequency (the frequency at which the swimmer updates its strategy) affects the found strategies. We find that higher update frequencies yield better performance.
\end{abstract}

\maketitle

\section{Introduction}
Many small swimming aquatic organisms, such as plankton, have sensing abilities that help them to navigate, locate prey and predators, and find potential mates~\cite{jiang2004hydrodynamics,kiorboe2008mechanistic,webster2009the,heuschele2014the}.
To do this efficiently, they need to be able to maneuver turbulent environments, in order to locate favourable flow regions, and to avoid predation.
The spatial distribution of simple organisms, such as swimming phytoplankton, can to a large extent be explained by passive mechanisms that do not require the organism to adapt its swimming to the environment~\cite{kessler1985hydrodynamic,guasto2012fluid,durham2013turbulence,zhan2014accumulation,gustavsson2016preferential,lovecchio2019chain}. However, certain species also respond actively to environmental cues~\cite{sengupta2017phytoplankton}.
Zooplankton have more advanced sensing abilities~\cite{yen1992mechanoreception,kioerboe1999hydrodynamic,fields2002mechanical,pecseli2016planktons}. Experiments show that they are able to navigate efficiently in turbulent flows by changing their swimming behavior in response to environmental cues~\cite{genin2005swimming,shang2008resisting,sidler2018counter,elmi2020the,elmi2022copepod,michalec2015turbulence,michalec2017zooplankton,adhikari2015simultaneous}.

Recently, reinforcement learning~\cite{colabrese2017flow,biferale2019zermelo,schneider2019optimal,alageshan2020machine,gunnarson2021learning,qiu2022active,xu2023long} as well as analytical approaches~\cite{biferale2019zermelo,liebchen2019optimal,moussaIder2021hydrodynamics,monthiller2022surfing,piro2023energetic,mousavi2024efficient} have been used to find optimal strategies for highly idealised
models of microswimmers in turbulence in different situations. Although these studies ignore many biological and hydrodynamical details that are believed to be significant, they nevertheless
offer a novel view on the long-standing and important problem of how plankton navigate in turbulent environments. This is important, because despite numerous experiments through the last decades, many fundamental questions of plankton navigation are still largely open.  There is no general understanding of which environmental cues are most important for different navigation tasks of aquatic microswimmers. Neither is it known how microswimmers may use environmental cues for efficient navigation.
The chief difficulty is that  empirical observations of the behaviour of microorganisms alone cannot give any information about optimal strategies unless the local flow configurations are precisely known, as well as which cues the organism measures.
In turbulence this is almost impossible to do unless one has a clear idea about which cues and strategies might be relevant. Here theoretical studies can help, provided that the models are based on simple, reasonable assumptions, and that the resulting strategies are efficient and robust. Most of the theoretical studies, however, rely on global information that is not available in the frame of reference of a swimmer~\cite{qiu2022navigation}. Moreover, most studies \cite{colabrese2017flow,gustavsson2017finding,qiu2022navigation,qiu2022active,monthiller2022surfing} consider very simple navigation problems, where the swimmer is supposed to cover a distance in a certain direction most efficiently or most quickly.
But in many cases for swimmers in nature or in applications, it may be more advantageous to search for, or avoid, certain flow regions.
One example is avoidance of high-strain regions of microswimmers in turbulent flows.
This is for example important for zooplankton because high turbulent strain masks the signal of an approaching predator, impeding the ability of the zooplankton to detect it early enough to escape~\cite{kioerboe1999hydrodynamic,jakobsen2001escape,buskey2002escape,kiorboe2008mechanistic}.

How can swimmers avoid regions of high fluid strain?
Ardeshiri {\em et al.} \citep{ardeshiri2016lagrangian} introduced a phenomenological model for the dynamics of copepods swimming by jumps in turbulence~\cite{ardeshiri2016lagrangian,ardeshiri2017copepods,kioerboe1999hydrodynamic,kioerboe1999predator}.
In this model, the copepods repeatedly jump head-first whenever the strain exceeds a threshold.
The dynamics of each jump follows the average jump behaviour observed in quiescent fluid.
This strategy helps the copepods to
avoid high strain regions to some extent, because the time spent in these regions is reduced. Since the strategy causes the copepods to move through the high-strain areas rather than avoiding them, it is plausible that the strategy is not optimal.
In Ref.~\cite{mousavi2024efficient}, we identified a simple yet highly efficient strategy allowing microswimmers to avoid high-strain regions altogether if they can measure the signum of local spatial strain gradients.
In principle, such signals can be detected using finite differences between sensory inputs from setae at different locations and orientations. However, plankton have likely evolved to detect simpler signals that correlate with the sign of local strain gradients.
Using a short-time expansion of the local, time-dependent spatial strain gradients  we could show that their signum is more important than the strain magnitude for strain avoidance.
This is plausible, because the strategy we found approaches gradient descent on the magnitude of the strain in the limit of  very large rotational swimming speeds and perfect sensing resolution.
For finite swimming and sensing abilities, the dynamics differs somewhat due to the inherent delay between sensing and turning.
Notably, the strategy is robust, maintaining excellent performance even for large sensing thresholds and in various flow configurations.

These results raise a number of important questions, which we address and answer in the following. First,  it is well known that the time horizon over which a behaviour is optimised can have a significant effect upon which strategy is found~\cite{recht2019tour}.
The standard approach for microswimmers~\cite{colabrese2017flow,biferale2019zermelo,schneider2019optimal,alageshan2020machine,gunnarson2021learning,qiu2022active,xu2023long} is to use a time horizon that is as large as possible, but the short-time expansion of Ref.~\cite{mousavi2024efficient} yields very good results.  \citet{monthiller2022surfing} used a similar short-time expansion to find optimal strategies for vertical navigation, and also found highly efficient strategies. It is surprising that the short-time expansion -- corresponding to a short time horizon -- works so well.
To find out why, we compare  the results of such short-time expansions for strain-avoiding microswimmers
with reinforcement-learning results, assuming that the microswimmers can/cannot measure the signum of strain gradients.
The reinforcement-learning simulations show that not only the time horizon matters, but also the frequency at which the swimmer measures the environment and updates its strategy.
We develop a new theory that explains how the optimal strategy depends on  the time horizon and upon the update frequency, for different signals. This allows us to conclude which signals are most efficient, depending on the time horizon.

These results are obtained for a highly idealised  model for the microswimmer. Therefore, the immediate significance for the behavioural science and evolutionary biology of swimming microorganisms is perhaps limited. Nevertheless,  the found strategies are robust and very efficient, and thus  potential candidate strategies in the ocean. However, it remains to be seen how precisely the swimmers can measure the signals used.
Last but not least, development of artificial microswimmers is rapidly progressing~\cite{cichos2020machine,tsang2020roads,muinos2021reinforcement,zou2022gait,mo2023challenges,rey2023light,amoudruz2024path,pradip2022deep, schrage2023ultrasound}, and it is expected that in the future, there will be engineered microswimmers with the ability to respond to signals from the fluid environment. It is therefore important to identify efficient navigation strategies for such swimmers for different types of signals.

\section{Methods}
\label{sec:model}

\begin{figure}
	\includegraphics[width=4.4cm,trim={0 0.2mm 0 0.5mm},clip]{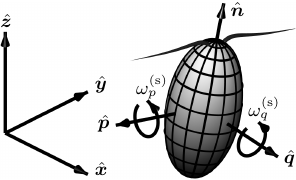}
	\caption{\label{fig:mechanism}
		Illustration of spheroidal microswimmer with swimming direction $\nhat$, direction of antennae $\phat$, and $\qhat=\nhat\times\phat$, in a fixed Cartesian frame of reference ($\hat{\ve x}$, $\hat{\ve y}$, $\hat{\ve z}$).
		Curved arrows denote positive directions of angular swimming velocities $\ve \omegasp$ and $\ve \omegasq$.
	}
\end{figure}
\subsection{Swimmer model}
We use a standard, highly idealised swimmer-model~\cite{kessler1985hydrodynamic,durham2009disruption,durham2013turbulence,qiu2022navigation,mousavi2024efficient} for a small spheroidal microswimmer with aspect ratio~$\lambda$:
\begin{subequations}
	\label{eq:eqm}
	\begin{align}
		\label{eq:eqm_x}
		\tfrac{\rm d}{{\rm d}t}{\ve x} & =\ve v\,,                                                                                                                    \\
		\tfrac{\rm d}{{\rm d}t}{\nhat} & =\ve\omega\cross\nhat
		\,,\hspace{0.5cm}
		\tfrac{\rm d}{{\rm d}t}{\phat}=\ve\omega\cross\phat
		\,,\hspace{0.5cm}
		\qhat=\nhat\cross\phat
		\,,                                                                                                                                                           \\
		\label{eq:eqm_v}
		\ve v(t) & =\ve u(\ve x,t)+\vs(t)\nhat\,,                                                                                               \\
		\ve\omega(t) & =\ve\Omega(\ve x,t)+\frac{\lambda^2-1}{\lambda^2+1}\nhat\cross(\ma S(\ve x,t)\nhat)+\omegas_p(t)\phat + \omegas_q(t)\qhat\,.
		\label{eq:eqm_omega}
	\end{align}
\end{subequations}
The body and antennae of the swimmer are assumed to be thin enough to have a negligible effect on the flow (one-way coupling).
The coordinate system $\nhat$, $\phat$, $\qhat$ rotates with the swimmer, $\nhat$ points forward to the head of the swimmer, $\phat$ is perpendicular to $\nhat$ and  lies in the direction of its antennae, and $\qhat$ is perpendicular to
$\nhat$ and  $\phat$ (Fig.~\ref{fig:mechanism}). The centre-of-mass velocity of the swimmer is denoted by  $\ve v$, and its angular velocity by $\ve\omega$. The flow velocity is  $\ve u(\ve x,t)$. The fluid-velocity gradients are $\ve\Omega$, half the flow vorticity, and
the strain-rate tensor $\ma S=(\ma A+\ma A^{\sf T})/2$ where $\ma A$ is the matrix of fluid-velocity gradients, with components $A_{ij}=\partial_ju_i$.

\subsection{Turbulent-flow models}
\label{sec:flow}
We use two different models for the turbulent-flow velocity $\ve u(\ve x,t)$ and its spatial derivatives: direct numerical simulation (DNS) and a statistical model.
The DNS are performed at Taylor-scale Reynolds number Re$_\lambda\approx 60$.
Moreover, we evaluate the short-time optimal strategies also for DNS at ${\rm Re}_\lambda\approx 418$ downloaded from the
Johns Hopkins University turbulence database~\cite{JohnsHopkins,JohnsHopkins2}.
The statistical model~\cite{gustavsson2016statistical,bec2024statistical}  represents the turbulent fluid velocities by Gaussian random functions with correlation
time $\tau_{\rm f}$, correlation length $\ellf$, and typical velocity $\uf$.
The ratio between the Eulerian time scale $\tauf$, and the Lagrangian time scale $\ellf/\uf$ defines a non-dimensional parameter of the model, the Kubo number $\ku = \uf \tauf / \ellf$~\cite{wilkinson2007unmixing}. The best agreement between the statistical model and DNS is found in the  limit of large but finite $\ku$~\cite{bec2024statistical}.
The limit of small Ku can be analysed with diffusion approximation and perturbation theory. Results obtained in this limit have yielded significant contributions to the understanding of the dynamics of inertial particles~\cite{duncan2005clustering}, and of  microswimmers~\cite{gustavsson2016preferential,borgnino2019alignment}
in turbulence. See Appendix~\ref{app:flow} for more details on the DNS and statistical models.

\subsection{Active control}
The terms with superscript (s) in Eqs.~(\ref{eq:eqm_v}) and (\ref{eq:eqm_omega}) allow the swimmer to actively swim and steer~\cite{qiu2022navigation,mousavi2024efficient,alageshan2020machine,biferale2019zermelo,gunnarson2021learning}.
The functions  $\vs(t)$, $\omegasp(t)$, and $\omegasq(t)$ parameterise possible actions in response to perceived changes to the environment of the swimmer.
Motivated by the behaviour of microorganisms in the turbulent ocean, we consider two different behaviours, cruising and jumping.
For the cruising swimmer, the swimming speed takes values $0\le\vs\le\vsmax$, where $\vsmax$ denotes the maximal swimming speed.
Rotational swimming occurs around both axes, $\phat$ and $\qhat$, with $-\omegasmax\le\omegasp,\omegasq\le\omegasmax$, where $\omegasmax$ is the maximal angular swimming speed for both $\omegasp$ and $\omegasq$.
Jumps, by contrast, are modeled as an instantaneous increase in the swimming speed, followed by an exponential decay with time scale $\tau_{\rm j}$
\begin{align}
	\vs(t)=
	\begin{array}{ll}
		2\log(10)\vsmax\exp\big(-\frac{t-t_{\rm j}}{\tauj}\big) & \mbox{for }t-t_{\rm j}<2\log(10)\tauj\,.
	\end{array}
	\label{eq:eqm_jump_velocity}
\end{align}
The jump is initiated at time $t_{\rm j}$, and the parameters in Eq.~(\ref{eq:eqm_jump_velocity}) are chosen so that $\vs(t)$ averaged over the total jump time $2\log(10)\tauj$ yields $0.99\vsmax$, so $\vsmax$ measures the average speed during a jump.
In our numerical simulations, we evaluate Eq.~(\ref{eq:eqm_jump_velocity}) at discrete time steps much smaller than the decay time scale $\tauj$, keeping the velocity constant in between time steps.
At the beginning of each jump, we allow the swimmer to rotate by either 0, 90, or -90 degrees around its $\qhat$-axis, respectively resulting in average angular velocities of $\omegasmax$, $0$, or $-\omegasmax$ during the jump, where $\omegasmax=\pi/(4\log(10)\tauj)$.

\subsection{Non-dimensional parameters}
Homogeneous isotropic turbulence is characterized by the Taylor-scale Reynolds number Re$_\lambda$, The fluctuations of the fluid velocity
in the statistical model are determined by the Kubo number $\ku$  mentioned above (see also Appendix~\ref{app:flow}). The non-dimensional parameters for the swimmer
are its aspect ratio  $\lambda$, as well as
\begin{align}
	\PhiK=\frac{\vsmax\tauK}{\etaK}\hspace{0.5cm}\mbox{ and }\hspace{0.5cm}\Xi=\omegasmax\tauK\,,
	\label{eq:Phi}
\end{align}
with $\tauK$ and $\etaK$ the Kolmogorov time and length scales, respectively.
This choice is natural for small-scale observables such as flow strain.
We obtain these scales in the stochastic model by matching covariances of first- and second-order flow gradients, see Appendix~\ref{app:flow}.
We assume slightly elongated swimmers with $\lambda=2$.
Typical values of $\PhiK$ and $\Xi$ vary widely depending on the swimming speed of the species, which ranges from \SIrange{1}{50}{\milli\meter\per\s}~\cite{svetlichny2020kinematic}, and the energy-dissipation rate per unit mass $\varepsilon$, which ranges from $10^{-4}\,\SI{}{\milli\meter\squared\per\s\cubed}$ in the deep ocean to $\SI{100}{\milli\meter\squared\per\s\cubed}$ in the upper ocean layer~\cite{yamazaki1996comparison,fuchs2016seascape}.
The resulting Kolmogorov time $\tauK$ decreases monotonically from 100 down to \SI{0.1}{\s}
and the Kolmogorov length $\eta$ decreases from \SI{10}{\milli\metre} to \SI{0.3}{\milli\metre}.
The smooth length scale ($\sim 10\eta$~\cite{pecseli2016planktons,bec2024statistical}) is much larger than typical planktonic microswimmers, except in extreme ocean turbulence.
For typical dissipation rates in the ocean, $\PhiK$ is much larger than unity, while in highly turbulent regions it is of the order unity.
Typical values of $\PhiK$ lie in the range $1$ to $100$.
Taking $\tauj=\SI{0.0088}{\s}$ for the jumping swimmer, in accordance with the experimental observations in Ref.~\cite{ardeshiri2016lagrangian}, leads to $\Xi\approx 39$ for $\tauK=\SI{1}{\s}$.
For cruising swimmers, showing less abrupt rotations, we use a smaller value, $\Xi=6.25$.
For  small Ku, the problem has to be non-dimensionalised in a different way~\cite{gustavsson2016statistical}. In this limit we use non-dimensional swimming and rotational swimming speeds $\vsmax\tauf/\ellf$ and $\omegasmax\tauf$.

\subsection{Reinforcement learning}
\label{sec:RL}
We  use reinforcement learning~\cite{sutton2018reinforcement,mehlig2021machine} to
search for efficient navigation strategies to avoid high-strain regions. This method uses trial and error to find optimal strategies for an agent interacting with an environment, and it has been used with great success for  microswimmers in turbulence~\cite{colabrese2017flow,biferale2019zermelo,schneider2019optimal,alageshan2020machine,gunnarson2021learning,muinos2021reinforcement,qiu2022navigation,qiu2022active,xu2023long}.
To connect with the standard terminology in reinforcement learning, we say that
the swimmer (agent) uses a strategy (policy) to respond (by taking certain actions) to environmental cues (states). The goal is to minimise the magnitude of strain, $\trSSt=\tr(\ma S(t)^2)\tauK^2$, it experiences.
We use two different prescriptions for sampling different states, which give same results. The control is either updated at regular time intervals $T_{\rm u}$, or after each time the state passes predefined levels. In the latter case, $\Tu$ is the average time between state updates.

When the state is updated, the swimmer obtains a reward
\begin{align}
	r_{i+1}=-\frac{1}{t_{i+1}-t_i}\int_{t_{i}}^{t_{i+1}}\!\!\!{\rm d}t\,\,\trSSt\,,
\end{align}
where $t_i$ and $t_{i+1}$ are the times at which two consecutive states $i$ and $i+1$ are encountered. Next, the action is updated according to the current strategy, and it is kept until the next state update.
The reinforcement learning algorithm iteratively maximizes the discounted expected future reward $R_{i+1}=\langle r_{i+1}+\gamma r_{i+2}+\gamma^2 r_{i+3}+\dots\rangle$,
where  $\langle \cdots\rangle$ denotes the average over all possible states and actions.
Furthermore,  $0\le\gamma<1$ sets the horizon determining the number of states  over which the reward is optimized.
The time horizon, $\sim\Tu/(1-\gamma)$, is kept much larger than both $T_{\rm u}$ and the smallest time scale of the flow.
Appendix~\ref{app:RL} summarises further details regarding the reinforcement learning algorithm and its parameters.
Below we consider cruising and jumping swimmers, with two corresponding sets of local states and actions, inspired by the most important signals identified in Ref~\cite{mousavi2024efficient}.

\begin{enumerate}
	\item {\em Strain magnitude}. In order to minimise the local strain $\trSSt = \tr(\ma{S}(t)^2) \tauK^2$, it is natural to use the signal $\sigma^2 = \tr(\ma S^2)\tauK^2$. We assume that the swimmer can distinguish 26 discrete levels of $\sigma^2$ that are more densely distributed for small values as shown in Table~\ref{tab:strategy}. This discretisation allows the reinforcement-learning algorithm to converge to good strategies in a reasonable time. Given a signal $\sigma^2$, we discretize the swimming actions into two levels: either no swimming, $\vs=0$, or to swim/jump with maximal speed, $\vs=\vsmax$.
	\item {\em Strain magnitude and derivatives.} In addition to the squared strain $\sigma^2$, we followed Ref.~\cite{mousavi2024efficient} and allowed the swimmer to measure the signum of derivatives of the squared strain in its reference frame, $X_\parallel=\nhat\cdot \ve \nabla\tr(\ma S^2)\tauK^2\ellf$, $X_\perp=\phat\cdot \ve \nabla\tr(\ma S^2)\tauK^2\ellf$ and $X^{(q)}_\perp=\qhat\cdot \ve \nabla\tr(\ma S^2)\tauK^2\ellf$.
We trained swimmers with $\sigma^2$ discretised into 5 levels for cruisers in the statistical model, and 26 levels for jumping swimmers in the DNS. Additionally, the signs of all $X_\parallel$, $X_\perp$ and $X^{(q)}_\perp$ were included in the state for the cruisers, while only the signs of $X_\parallel$ and $X_\perp$ were included for the jumpers. In addition to propulsion/jump speed of either 0 or $\vsmax$ in their instantaneous direction, we include actions of rotational swimming. For the cruisers, we assume that the angular swimming velocities in Eq.~(\ref{eq:eqm_omega}), $\omegasp$ and $\omegasq$, each can take either of the values $\{-\omegasmax,0,\omegasmax\}$, while for the jumping swimmers, the swimmer prior to each jump either keeps its current orientation, or instantaneously rotates by an angle of $\pm\pi/2$ around its $\qhat$ axis.
\end{enumerate}

\begin{table}
	\includegraphics[width=0.7\textwidth]{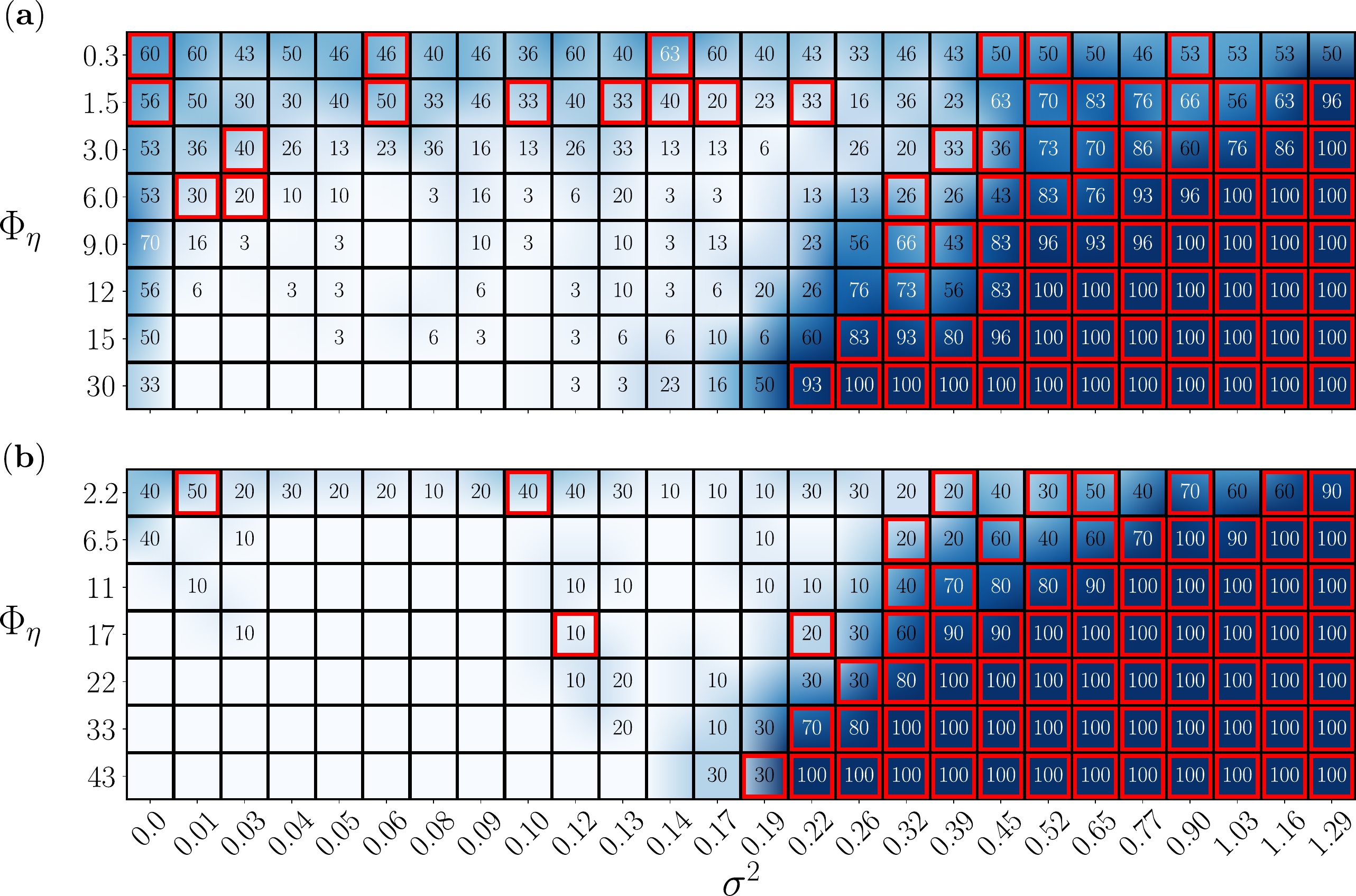}
	\caption{\label{tab:strategy}
		Strategies resulting from reinforcement learning for ({\bf a}) cruising swimmers in the statistical model and ({\bf b}) jumping swimmers in DNS for different average swimming speed $\PhiK$.
		The values on the horizontal axis represents lower limits of the discretized state of squared strain, $\sigma^2$.
		The cell numbers indicate the percentage of strategies that takes the action to swim/jump for a given state, color coded from white (0\%) to clear blue (100\%).
		For each $\PhiK$, the best strategy found is highlighted with a red frame for states where it swims/jumps.
		Each value of $\PhiK$ show data based on ({\bf a}) 30 training sessions and ({\bf b}) 10 sessions.
	}
\end{table}

\subsection{Short-time expansion}
The reinforcement-learning study presented here is motivated by the results of Ref.~\cite{mousavi2024efficient}, where a short-time expansion of Lagrangian signals was used to find optimal strategies. A related short-time expansion was used to find navigation strategies for vertical migration of microswimmers~\cite{monthiller2022surfing}.  In the following we briefly outline the method~\cite{mousavi2024efficient}.
Consider first the case where the swimmer measures only the magnitude of the strain.
We assume large $\PhiK$ and $\omegas=0$ to approximate $\ve x_t=\ve x_0+\vs\ve \nhat_0t$. We expand $\trSS(x_t,t)=\tr(\ma S^2(x_t,t))\tauK^2$ sampled after a state update at time $t=0$ to second order in small $\vs t/\tauK$,
\begin{align}
	\trSSarg{\ve x_t,t}=\trSSarg{\ve x_0,t}+\vs t\nhat_0\cdot\ve\nabla\trSSarg{\ve x_0,t}+\frac{1}{2}[\vs t]^2(\nhat_0\cdot\ve\nabla)^2\trSSarg{\ve x_0,t}\,.
\end{align}
We average this expression conditional on $\sigma^2=\tr(\ma S^2(\ve x_0,0))\tauK^2$, assuming Gaussian distributed fluid gradients and uniformly distributed orientations $\nhat_0$ at the state update.
For the statistical model described in Section \ref{sec:flow}, we can evaluate the resulting expression explicitly. In spatial dimension $d$ we find:
\begin{align}
	\langle\trSSarg{\ve x_t,t}|\sigma^2\rangle=\frac{1}{2}+\Big(\sigma^2-\frac{1}{2}\Big)e^{-2t/\tauf}\Big(1-\frac{d+4}{d}\Big[\frac{\vs t}{\ellf}\Big]^2\Big)\,.
	\label{eq:TrSSqrConditionalTrSSqrSmallR}
\end{align}
The first term on the right-hand side is the average obtained for a uniform spatial distribution of particles, $\sigma^2_{\rm c}=\langle\tr(\ma S^2)\tauK^2\rangle=1/2$.
The second term is a correction due to the initial strain differing from this average.
The optimal swimming velocity that minimizes Eq.~(\ref{eq:TrSSqrConditionalTrSSqrSmallR}) is to swim with maximal speed when $\sigma^2>\sigma^2_{\rm c}$, and to not swim at all otherwise:
\begin{align}
	\vsopt(\sigma^2)=\left\{
	\begin{array}{ll}
		\vsmax & \mbox{if }\sigma^2>\sigma^2_{\rm c}\cr
		0      & \mbox{otherwise}
	\end{array}
	\right.\,,\hspace{0.5cm}\mbox{with }\sigma^2_{\rm c}=\frac{1}{2}\,.
	\label{eq:optimal_policy_TrSSqr}
\end{align}

Now consider the second case, where the swimmer can measure strain gradients as well. This was considered in  Ref.~\cite{mousavi2024efficient}, where a short-time expansion was used
to show that  strain gradients are the most important signals for avoiding strain at short times. The resulting strategy for choosing the swimming speed $\vs$ and rotational swimming $\omegasp(t)$ and $\omegasq(t)$ was found to be
\begin{align}
	\vs_{\rm opt}(t) & =\left\{
	\begin{array}{ll}
		\vsmax & \mbox{if }X_\parallel<0\cr
		0      & \mbox{otherwise}
	\end{array}\right.\,,\hspace{0.5cm}
	\omegas_{p,{\rm opt}}(t) =\omegasmax\text{sign}(X^{(q)}_\perp)\,,\hspace{0.5cm}
	\omegas_{q,{\rm opt}}(t) =-\omegasmax \text{sign}(X_\perp)\,.
	\label{eq:optimal_policy}
\end{align}
Below we compare both strategies, Eqs.~(\ref{eq:optimal_policy_TrSSqr}) and (\ref{eq:optimal_policy}), to the numerical results found by reinforcement learning.
\begin{figure}
	\includegraphics[width=0.8\textwidth]{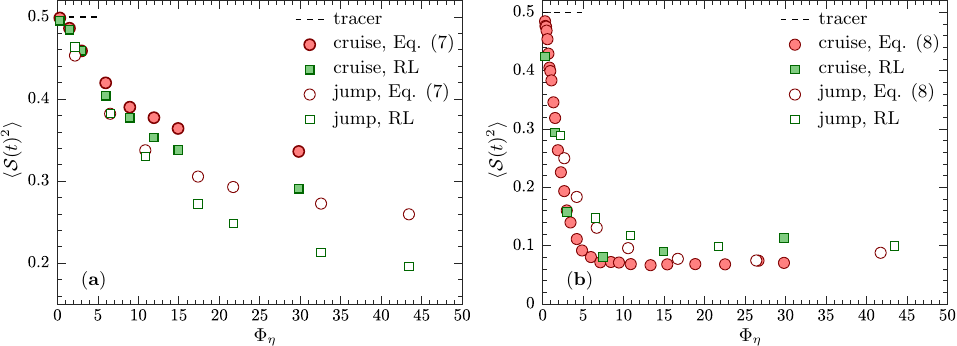}
	\caption{\label{fig:RL_performance}
		Comparison of numerical results for the average $\trSS$ against average swimming speed $\PhiK$ for swimming strategies ({\bf a}) excluding and ({\bf b}) including strain gradients as signal.
		({\bf a}) Results for reinforcement learning (RL) in Table~\ref{tab:strategy} for cruising swimmers in the statistical model (green, $\blacksquare$) and jumping ones in DNS ($\Box$), and for the analytical strategy in Eq.~(\ref{eq:optimal_policy_TrSSqr}) for cruising (red, $\bullet$) and jumping ($\circ$).
			{(\bf b)} Results for RL in Table~\ref{tab:strategy_general} for cruising (green, $\blacksquare$) and jumping ($\Box$) swimmers, and for Eq.~(\ref{eq:optimal_policy}) evaluated for cruising swimmers (red, $\bullet$) and jumping swimmers disregarding the signal $X^{(q)}_\perp$ ($\circ$).
		Parameters: $\Xi=6.25$ for cruising swimmers and $\Xi=39$ for jumping ones, and $\lambda=2$.
	}
\end{figure}

\begin{table}
	\begin{overpic}[width=\textwidth]{{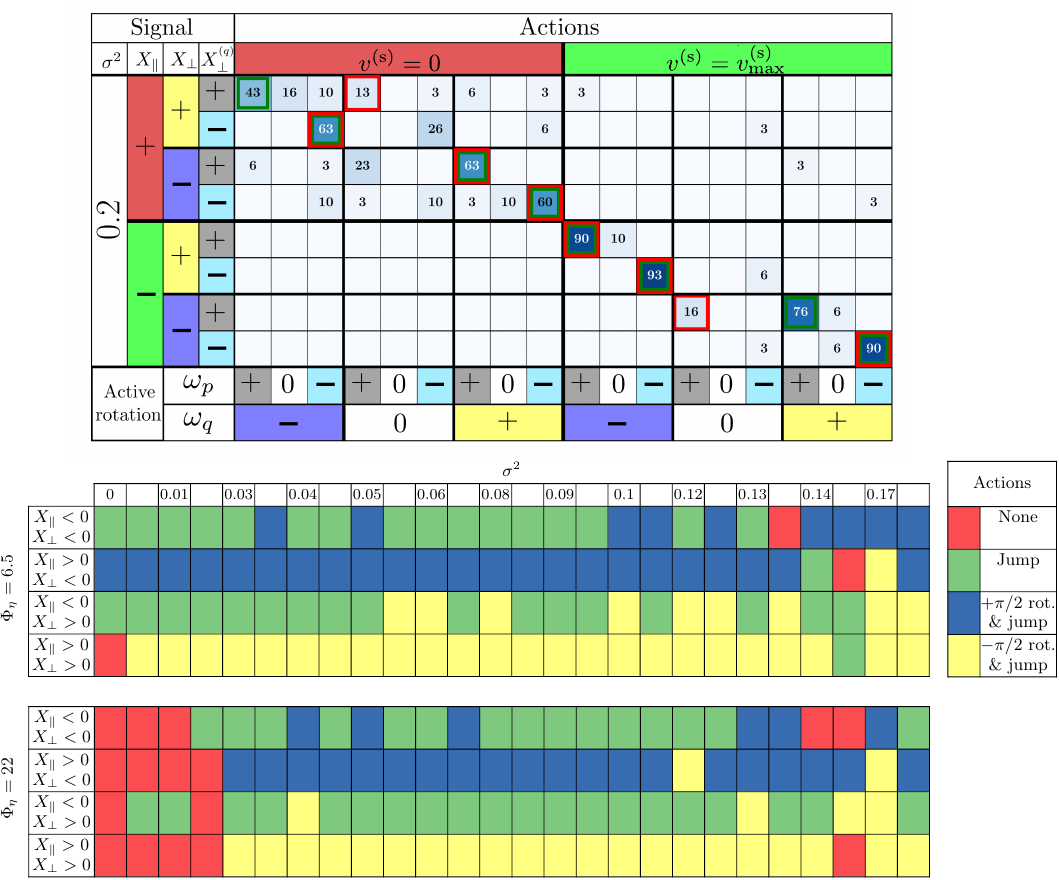}}
	\end{overpic}
	\caption{\label{tab:strategy_general}
		Strategies resulting from reinforcement learning for ({\bf a}) cruising swimmers with $\PhiK=7.5$ and $\Xi=6.25$, and ({\bf b}) jumping swimmers with $\PhiK=6.5$ and $\PhiK=22$, and $\Xi=39$.
		({\bf a}) Simplified strategy, showing only the state $0.1<\sigma^2\le 0.2$, using a threshold $X_{\rm c}$ for $X_\parallel$, $X_\perp$, and $X^{(q)}_\perp$ (see text).
		Numbers give percentage of 30 trained strategies taking an action (column) in a state (row).
		Red frames show the best found strategy, and green frames show Eq.~(\ref{eq:optimal_policy}).
		({\bf b}) States are given by $\sigma^2$ (columns) and the signs of $X_\parallel$ and $X_\perp$ (rows).
		The action taken in each state for the best strategy found in training is color coded to do nothing (red), jump forward (green), rotate $\pm90\degree$ around $\qhat$ and then jump (blue/yellow).
	}
\end{table}

\section{Numerical results}
\subsection{Navigation based on strain magnitude}
\label{sec:RLstrain}
The reinforcement-learning strategies for different values of $\PhiK$ are shown in Table~\ref{tab:strategy} for the case of cruising swimmers in the statistical model and for jumping swimmers in DNS.
We see that the strategies are very similar for the two cases.
For small $\PhiK$, the swimmer does not learn much and either swims or stops with approximately equal probability.
For $\PhiK$ larger than unity, the training converges to more clearcut strategies.
We see that the swimmer learns to swim/jump when encountering large strain rates, and it tends to stop swimming for small strain rates.
The threshold value between swimming and stopping depends on the swimming speed $\PhiK$, with a smaller threshold value for larger $\PhiK$.
Additionally, for cruising swimmers we tested continuous signal $\sigma^2$ and action $\vs$ in deep reinforcement learning, as well as adding rotational swimming to Q-learning, all giving essentially the same strategy: swim above a strain threshold and not swim below it (see Appendix~\ref{app:DQN}).

Fig.~\ref{fig:RL_performance}({\bf a}) shows the average squared strain evaluated following the best strategies in Table~\ref{tab:strategy} for cruising (green filled squares) and jumping (empty squares) swimmers.
The averages decrease monotonically with the swimming speed $\PhiK$ and is substantially smaller than the average of tracer particles, unless $\PhiK$ is too small.
For a given $\PhiK$, the average strain is lower for the jumping swimmer than the cruising one.
The difference is mainly due to different flow statistics in the DNS and the model. Simulations of cruising swimmers in DNS have approximately the same performance as the jumpiong ones, see Fig.~\ref{fig:cruiser}({\bf a}) in Appendix~\ref{app:cruiser}.
In summary, the two approaches exhibit similar trends. They yield similar performance for $\PhiK\sim 1$, but the strategy based on reinforcement learning performs better for large~$\PhiK$.

\subsection{Navigation based on squared strain and its derivatives}
\label{sec:RLstrain_gradients}
Examples of the best reinforcement-learning strategies for swimmers sensing squared strain and its derivatives are shown in Table~\ref{tab:strategy_general}.
Table~\ref{tab:strategy_general}({\bf a}) shows an example for cruising swimmers with $\PhiK=7.5$ restricted to the state with $\sigma^2=0.2$, and Table~\ref{tab:strategy_general}({\bf b}) shows two examples for jumping swimmers with $\PhiK=6.5$ and $22$.
For cruising swimmers, the majority of strategies and the best strategy essentially coincides with Eq.~(\ref{eq:optimal_policy}), which is highlighted in green.
There are deviations, but a comparison shows that Eq.~(\ref{eq:optimal_policy}) gives a slightly better performance, suggesting that these deviations stem from the learning getting stuck at local optima.
The strategy is qualitatively the same for the other $\sigma^2$ states.
It also remains the same if the signals $X_i$ are instead discretized in three levels symmetrically distributed around zero separated at $\pm 1$.
These results are consistent with Ref.~\cite{mousavi2024efficient}, where it was found that the performance of Eq.~(\ref{eq:optimal_policy}) does not change notably by introducing a sensing threshold.
The jumping swimmer learns a similar strategy.
The trend is to always jump.
When $X_\parallel<0$, the jump is mainly forward, allowing the swimmer to decrease its expected strain, similar to the behavior of the cruising swimmer.
When $X_\parallel>0$, it still makes a jump, but with a $\pm\pi/2$ rotation, depending on the sign of $X_\perp$, to bias its resulting orientation in a direction where $\trSS$ decreases.
The jump directions are the same as for the cruiser and Eq.~(\ref{eq:optimal_policy}).
For the case $\PhiK=22$ and $\sigma^2<0.03$, there is a trend to not jump.
This is expected because regions with $\sigma^2\ll 1$ have small volume, meaning that a jump will most likely overshoot, causing the swimmer to end up in a region of higher strain.

Fig.~\ref{fig:RL_performance}({\bf b}) compares the performance of the reinforcement learning and Eq.~(\ref{eq:optimal_policy}) for cruising and jumping swimmers.
The performance is qualitatively the same for all cases, with a sharp drop from the limit of tracer particles at small swimming speeds to values $\langle\trSSt\rangle\sim 0.1$ for swimming speeds larger than order unity.
This is significantly lower than the results based only on strain in Fig.~\ref{fig:RL_performance}({\bf a}), showing that gradients of $\sigma^2$ are more efficient signals for avoiding high strain regions.
In contrast to the strategy based on $\sigma^2$, jumping swimmers do not perform better than cruising ones.
This is also the case for cruising swimmers in DNS, see Fig.~\ref{fig:cruiser}(b) in Appendix~\ref{app:cruiser}, showing that the cruising swimmer dynamics approximates jumping swimmer dynamics well.

\section{Approximate theory for optimization on different time horizons}
\label{sec:analytical}
The comparison between the results from reinforcement learning optimized on a large time horizon, and the theory derived at a short time horizon in Fig.~\ref{fig:RL_performance} shows that there is no essential difference when the gradient of squared strain is used as signal.
However, when the squared strain is used as signal, reinforcement learning gives a better strategy.
To explain this difference, and to more generally investigate the importance of the time horizon, we develop an approximate theory for how the optimal solution depends on the optimization horizon.
The theory can be used to identify the most important signals and gives a lowest-order approximation for suitable swimming strategies based on selected flow signals.

\subsection{Theory}
The theory examines a scenario in which a spherical swimmer remains inactive for a duration exceeding the characteristic time scale of the flow. The swimmer then makes a single measurement of the flow at $t=0$ and subsequently swim with constant velocity $\vs(t)=\vs$ and angular velocity $\omegas(t)=\omegas$ for a time period $T_{\rm p}$.
This scenario allows to analytically predict how the average strain, conditional on the values of different initial flow signals, evolves in time for all choices of $\vs$ and $\omegas$.
The truly optimal swimming strategy in this scenario is obtained by selecting $\vs=\vsopt$ and $\omegas=\omegasopt$ that minimize the conditional strain time averaged up to the prediction time $\Tp$.
To calculate this average, we make an analytical expansion of the dynamics.
For simplicity, we consider the statistical model in two spatial dimensions.
Since cruising and jumping swimmers with the same average swimming speed have the same performance in our DNS (Figs.~\ref{fig:RL_performance} and~\ref{fig:cruiser}), we focus only on cruising swimmers.
A lowest-order approximation to the solution of the swimmer dynamics in Eqs.~(\ref{eq:eqm}) is obtained by
\begin{subequations}
\label{eq:det}
	\begin{align}
		\label{eq:xdet}
		{\ve x}^{(\rm d)}_t & =\ve x_0+\frac{\vs}{\omegasq}\left(\sin(\omegasq t)\nhat_0 + (1-\cos(\omegasq t))\phat_0\right) \\
		\nhat^{(\rm d)}_t & = \cos(\omegasq t)\nhat_0 + \sin(\omegasq t)\phat_0\,,
		\label{eq:ndet}
	\end{align}
\end{subequations}
where $\ve x_0$, $\nhat_0$, and $\phat_0=\hat{\ve z}\times\nhat_0$ are the initial position and orientation vectors.
This solution is valid if the swimming velocity is much larger than the velocity of turbulent fluctuations, as is usually the case for zooplankton in the ocean, or if the Kubo number is small.
In the limit $\omegasq\to 0$, the trajectory simplifies to ${\ve x}^{(\rm d)}_t=\ve x_0+\vs\nhat_0 t$, as expected for swimmers without rotational swimming.

In the scenario outlined above, the components of the flow and its derivatives are initially Gaussian distributed and the swimmer orientations are uniformly distributed, independent from the flow.
We evaluate the average strain along the trajectory (\ref{eq:xdet}) conditional on the initial strain tensor $\ma S_{0}$, strain tensor gradient $\ve\nabla\ma S_{0}$ (with components $\partial_kS_{0,ij}$), projected on the initial orientation of the swimmer, parameterized by $\nhat_{0}$ and $\phat_{0}$.
We do not condition on the initial flow velocity and vorticity, because these are not directly measurable in the frame of the swimmer using setae only.
Moreover, we have used that all second-order derivatives of the flow can be expressed in terms of $\ve\nabla\ma S_{0}$, meaning that the signal $\ve\nabla\ma O_{0}$ is redundant, and therefore not included.
Finally, we have neglected third-order and higher derivatives of the flow.
For Gaussian distributed flow components, we have
\begin{align}
	\langle\trSSt|\ma S_0,\ve\nabla\ma S_0\rangle=\sum_{i=1}^d\sum_{j=1}^d\langle S_{ij}(t)^2|\ma S_0,\ve\nabla\ma S_0\rangle
	=\sum_{i=1}^d\sum_{j=1}^d[\langle S_{ij}(t)^2\rangle-\ve B^{\sf T}(S_{ij}(t))\ma C^{-1}\ve B(S_{ij}(t))+(\ve B^{\sf T}(S_{ij}(t))\ma C^{-1}\ve F)^2]
\end{align}
with
\begin{align}
	\ve B(S_{ij})=\begin{pmatrix}
		              \langle S_{ij}F_1\rangle\cr
		              \vdots\cr
		              \langle S_{ij}F_N\rangle
	              \end{pmatrix}
	\,,\;
	\ma C=\begin{pmatrix}
		      \langle F_1^2\rangle  & \cdots & \langle F_1F_N\rangle \cr
		      \vdots                & \ddots & \vdots\cr
		      \langle F_1F_N\rangle & \cdots & \langle F_N^2\rangle
	      \end{pmatrix}
	\,,\;
	\ve F=\begin{pmatrix}
		      F_1\cr
		      \vdots\cr
		      F_N
	      \end{pmatrix}
\end{align}
where $F_1$,\dots,$F_N$ enumerate all components of $\ma S_0$ and $\ve\nabla\ma S_0$,
$\ma S(t)$ is evaluated along a deterministic trajectory (\ref{eq:xdet}), and the averages in $\ve B$ and $\ma C$ are explicitly known in the statistical model.

Using the isotropic, Gaussian distributed correlation function (\ref{eq:SMcorr}) for the flow in two spatial dimensions, we obtain
\begin{align}
	 & \langle\trSSt|\sigma_{\alpha\beta}=S_{0,\alpha\beta}\tauK^2,\xi_{\alpha\beta\gamma}=\partial_\gamma S_{0,\alpha\beta}\tauK^2\ellf\rangle=
	\nonumber\\
	 & \hspace{0.5cm}
	\frac{1}{2}+e^{-2t/\tauf-R^2}\bigg[
	C_1\Big(\sigma_{ij}\sigma_{ij}-\frac{1}{2}\Big)
	-C_2\Big(\sigma_{rr}^2-\frac{1}{8}\Big)
	+C_3\Big(\xi_{ijk}\xi_{ijk}-3\Big)
	-C_4\Big(\xi_{ijk}\xi_{ikj}-\frac{3}{2}\Big)
	+C_5\Big(\xi_{ijr}\xi_{ijr}-\frac{3}{2}\Big)
	\nonumber\\&\hspace{2cm}
	+C_{6}\Big(\xi_{irr}\xi_{rri} - \frac{3}{8}\Big)
	-C_{7}\Big(\xi_{rrr}\xi_{rrr} - \frac{3}{8}\Big)
	+C_{8} \sigma_{ij}\xi_{ijr}
	+C_{9} \sigma_{ij}\xi_{rij}
	+C_{10} \sigma_{rr}\xi_{iri}
	-C_{11} \sigma_{rr}\xi_{rrr}
	\bigg]\,.
	\label{eq:TrSSqrConditionalGeneral}
\end{align}
Here indices $i,j,k$ are summed over, the subscript $r$ denotes contraction with $\hat{\ve R}(t)=\ve R(t)/|\ve R(t)|$, and $R=|\ve R(t)|$, where $\ve R(t)=(\ve x_t^{\rm (d)}-\ve x_0)/\ellf$ with initial orientation $\hat{\ve R}(0)=\nhat_0$.
The coefficients $C_i$ are polynomials in $R$ with sign chosen positive for small $R$:
\begin{align}
	\begin{split}
		C_1   & =(1-\tfrac{R^2}{2})^4\,,\;
		C_2=R^4(1-R^2+\tfrac{R^4}{8})\,,\;
		C_3=\tfrac{R^4}{36}(6-R^2)\,,\;
		C_4=\tfrac{R^4}{9}(3-R^2)\,,\;
		C_5=\tfrac{R^2}{144}(144-192R^2+84^4-16^6+R^8)\,,\;
		\nonumber                                   \\
		C_{6} & =\tfrac{R^4}{18}(24-12R^2+R^4)\,,\;
		C_{7}=\tfrac{R^6}{72}(24-12R^2+R^4)\,,\;
		C_{8}=\tfrac{R}{24}(2-R^2)^2(12-10R^2+R^4)\,,\;
		\\
		C_{9} & =\tfrac{R^3}{6}(4-4R^2+R^4)\,,\;
		C_{10}=\tfrac{R^3}{6}(8-8R^2+R^4)\,,\;
		C_{11}=\tfrac{R^5}{12}(16-10R^2+R^4)\,.
		\nonumber
	\end{split}
\end{align}
For each term containing $C_i$, the ensemble average over the flow has been subtracted, meaning that each such term averages to zero.

For short times, $R$ is small and the dominant contribution to Eq.~(\ref{eq:TrSSqrConditionalGeneral}) is given by the $C_1$ and $C_8$ terms.
By averaging over the strain gradients $\xi_{\alpha\beta\gamma}$, neglecting rotational swimming and expanding to order $t^2$ gives
\begin{align}
	 & \langle\trSSt|\sigma_{\alpha\beta}=S_{0,\alpha\beta}\tauK^2\rangle
	=\frac{1}{2}+\Big(\sigma_{ij}\sigma_{ij}-\frac{1}{2}\Big)e^{-2t/\tauf}\Big(1-3\Big[\frac{\vs t}{\ellf}\Big]^2\Big)\,,
	\label{eq:TrSSqrConditionalTrSSqrSmallt}
\end{align}
i.e. Eq.~(\ref{eq:TrSSqrConditionalTrSSqrSmallR}) in $d=2$ spatial dimensions.
Similarly, expanding Eq.~(\ref{eq:TrSSqrConditionalGeneral}) for small $t$, keeping terms to zeroth order in $t$, but also the dominant contributions containing the controls $\vs$ and $\omegasq$, gives (in dimensional units)
\begin{align}
	\langle\tr(\ma S(t)^2)|S_{0,\alpha\beta},\partial_\gamma S_{0,\alpha\beta}\rangle & \sim
	\tr(\ma S_0^2)+\vs t\nhat_0\cdot\ve\nabla\tr(\ma S_0^2) + \frac{\vs\omegasq}{2}t^2\phat_0\cdot\ve\nabla\tr(\ma S_0^2)\,.
	\label{eq:TrSSqrConditionalSmallt}
\end{align}
This result is equivalent to a two-dimensional version of the short-time expansion in Ref.~\cite{mousavi2024efficient}.
Equations~(\ref{eq:TrSSqrConditionalTrSSqrSmallt}) and~(\ref{eq:TrSSqrConditionalSmallt}) are minimized by the control in Eqs.~(\ref{eq:optimal_policy_TrSSqr}) and (\ref{eq:optimal_policy}), respectively.

Since Eq.~(\ref{eq:TrSSqrConditionalGeneral}) predicts the evolution of the conditional strain on arbitrary times, this allows to move away from the short-time expansion, allowing to formulate the strategy that is optimal on an arbitrary time horizon.
We obtain the average squared strain conditional on different combinations of strain and strain gradients by averaging Eq.~(\ref{eq:TrSSqrConditionalGeneral}) using the distribution conditional on the desired signals, assuming that the initial orientation is uniformly distributed.
Starting from an initial flow signal IC at time $t$, we evaluate the time average of the conditional average during the prediction time interval $\Tp$
\begin{align}
	\overline{\langle\trSSt|IC\rangle}(\Tp)=\frac{1}{\Tp}\int_{0}^{\Tp}{\rm d}t\langle\trSSt|IC\rangle\,.
	\label{eq:TimeAverageStrain}
\end{align}
Below we analyze the optimal strategies that minimize Eq.~(\ref{eq:TimeAverageStrain}) for different signals. We constrain velocities by $\vsmax=\ellf/\tauf$ and $\omegasmax=5\tauf$.
For larger $\vsmax$, contributions from the tails of the spatial correlation function start to matter. These tails are different in the single-scale statistical model and in turbulence with an inertial range.

\subsection{Application to squared strain}
\label{sec:ApplicationTrSSqr}

\begin{figure}
	\begin{overpic}[width=0.7\textwidth]{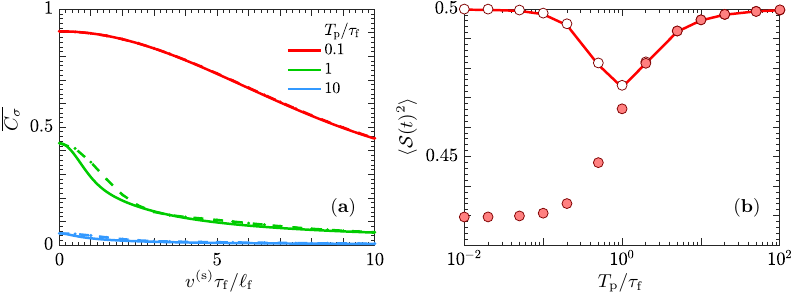}
	\end{overpic}
	\caption{
		\label{fig:GaussConditionalStrain}
		({\bf a}) Time averaged coefficient $\overline{C_\sigma}(\Tp)$ (see text) against swimming velocity for different prediction horizons $\Tp$ with rotational swimming ($\omegas\tauf=5$, dashed lines) and without ($\omegas=0$, solid lines).
		({\bf b}) Average squared strain, $\langle\trSSt\rangle=\langle\tr(\ma S(t)^2)\rangle\tauK^2$ against $\Tp$, following the optimal strategy.
		The solid line shows an analytical evaluation of the predicted time averaged strain (see text).
		Markers show the time average from numerical simulations of swimmers following Eq.~(\ref{eq:optimal_policy_TrSSqr}).
		Results are obtained by either choosing the initial position randomly (red, $\circ$), or by sequentially measuring the signal and updating the control each time interval $\Tu=\Tp$ (red, $\bullet$).
		Parameters: $\vsmax\tauf/\ellf=1$, $\omegasmax\tauf=5$, $\lambda=1$, and $\ku=0.1$.
	}
\end{figure}

Averaging Eq.~(\ref{eq:TrSSqrConditionalGeneral}) conditional on a certain value of $\sigma^2=\tr(\ma S_0^2)\tauK^2$ gives
\begin{align}
	\langle\trSSt|\sigma^2\rangle = \frac{1}{2} + e^{-2t/\tauf-R^2}\frac{1}{32}(32-64R^2+40R^4-8R^6+R^8)\Big(\sigma^2-\frac{1}{2}\Big)\,.
	\label{eq:TrSSqrConditionalTrSSqr}
\end{align}
The first term is the average of uniformly distributed particles, $\sigma^2_{\rm c}=\langle\tr(\ma S^2)\tauK^2\rangle=1/2$. The second term describes the temporal relaxation of the initial squared strain $\sigma^2$ towards this value. It is given by a coefficient $C_{\sigma}$ multiplying $\sigma^2-\frac{1}{2}$.
Fig.~\ref{fig:GaussConditionalStrain}({\bf a}) shows the time average $\overline{C_{\sigma}}(\Tp)=\tfrac{1}{\Tp}\int_0^{\Tp}{\rm d}t\,C_\sigma$ against the constant swimming velocity $\vs$.
The coefficient is positive and monotonically decreasing with $\vs$ for different prediction horizons $\Tp$.
This implies that the optimal swimming velocity to minimize the time average $\overline{\langle\trSSt|\sigma^2\rangle}(\Tp)=\sigma^2_{\rm c}+\overline{C_{\sigma}}(\Tp)(\sigma^2-\sigma^2_{\rm c})$ in Eq.~(\ref{eq:TrSSqrConditionalTrSSqr}) is identical to the strategy in Eq.~(\ref{eq:optimal_policy_TrSSqr}).
Rotational swimming does not contribute much [dashed lines in Fig.~\ref{fig:GaussConditionalStrain}({\bf a})], and it mainly slows down the decorrelation.
We therefore conclude that the optimal rotational swimming is $\omegasopt(\sigma^2)=0$ when navigating using $\sigma^2$ as the signal.
The form of the strategy in Eq.~(\ref{eq:optimal_policy_TrSSqr}) is independent of the prediction horizon $\Tp$, but the relative decrease in $\overline{\langle\trSSt|\sigma^2\rangle}(\Tp)$ compared to tracer particles is largest around $\Tp\sim\tauf$.

The performances of the optimal strategy (\ref{eq:optimal_policy_TrSSqr}) is shown in Fig.~\ref{fig:GaussConditionalStrain}({\bf b}).
The solid line shows the time average of the theoretical prediction in Eq.~(\ref{eq:TrSSqrConditionalTrSSqr}) using time-independent $\vs=\vsopt(\sigma)$ and $\omegas=\omegasopt(\sigma)$, and additionally averaged over $\sigma^2$, assuming Gaussian distributed strain components.
Hollow markers show the time average of $\trSS$ obtained by numerical simulations of Eq.~(\ref{eq:eqm}), starting from a random position and following the optimal strategy (\ref{eq:optimal_policy_TrSSqr}) in the statistical model over the prediction time $\Tp$ .
The simulations agree well with the theoretical predictions.
There is an optimum around $\Tp=\tauf$, showing that the strategy efficiently exploit flow correlations on this time scale. For larger times, the flow decorrelates from the initial condition, meaning that prediction is no longer possible.

For microswimmers that sense their environment and continuously adjust their behavior, we consider an additional scenario where the swimmer updates $\vs$ and $\omegasq$ at regular time intervals $\Tu$. This is the setup we use in our reinforcement learning.
At each update, the flow signal $\sigma^2$ is evaluated at the current position, giving new values for $\vs=\vsopt(\sigma)$ and $\omegas=\omegasopt(\sigma)$, determined by the optimal strategy derived for the original scenario.
The history of choices of $\vs$ and $\omegas$ may influence preferential alignment and sampling of strain, causing the flow distribution at update times to deviate from the Gaussian distribution assumed in the derivation of the theory.
This effect becomes stronger as $\Tu$ decreases.

Filled markers in Fig.~\ref{fig:GaussConditionalStrain}({\bf b}) show simulation results for the average squared strain sampled continuously using an update time $\Tu=\Tp$. The results agree with the theory when $\Tp$ is larger than $\tauf$, but for $\Tp$ smaller than $\tauf$, the values are significantly lower than those predicted by theory.
This is expected because the strain distribution becomes increasingly non-Gaussian in this limit and skewed towards smaller $\trSS$ when following the optimal strategy.
The non-Gaussian distribution of strain at update times enhances the performance of the optimal control derived from a Gaussian distribution, showing monotonical improvement as $\Tu=\Tp$ decreases.

\subsection{Application to gradients of squared strain}
\label{sec:ApplicationGradient}
To investigate how gradients of strain can be utilized for optimizing the sampled squared strain on different time horizons, we first consider either the longitudinal or transversal gradients, $X_\parallel=\nhat_0\cdot\ve\nabla\tr(\ma S^2_0)\tauK^2\ellf$ or $X_\perp=\phat_0\cdot\ve\nabla\tr(\ma S^2_0)\tauK^2\ellf$ as signals.
Moreover, we evaluate the optimal strategy based on all components of strain and strain gradients, $\ma S_0\tauK^2$ and $\ve\nabla\ma S_0\tauK^2\ellf$ in the $\nhat_0$, $\phat_0$ basis.

\begin{figure}
	\begin{overpic}[width=\textwidth]{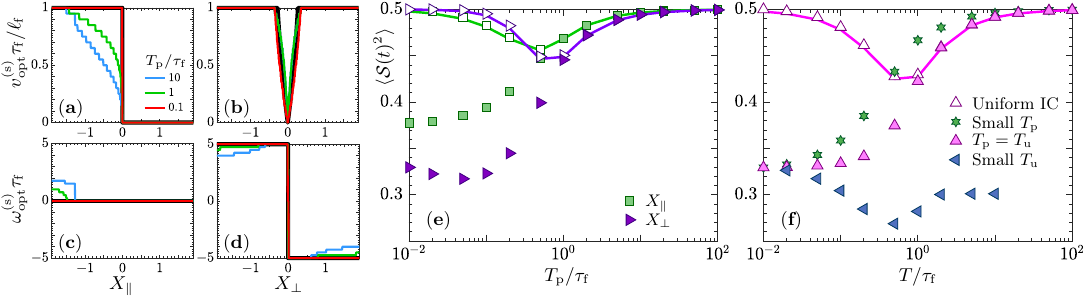}
	\end{overpic}
	\caption{
	\label{fig:GaussConditional}
	Navigation based on derivatives of squared strain.
	({\bf a}--{\bf d}) Optimal strategy for choosing ({\bf a},{\bf b}) $\vs$ and ({\bf c},{\bf d}) $\omegas$ based on the signals ({\bf a},{\bf c}) $X_\parallel$ and ({\bf b},{\bf d}) $X_\perp$ for different prediction horizons $\Tp$ (solid lines).
	Data is obtained by evaluating the time average of Eqs.~(\ref{eq:TrSSqrConditionalX}) and (\ref{eq:TrSSqrConditionalY}) for a discrete set of $\vs$ and $\omegasq$, and choosing $\vsopt$ and $\omegasopt$ that gives the smallest average squared strain.
	Thick black lines show the small-$\Tp$ limit in Eqs.~(\ref{eq:optimal_policy_X}) and (\ref{eq:optimal_policy_Y}).
	({\bf e}) Same as Fig.~\ref{fig:GaussConditionalStrain}({\bf b}) for the signals $X_\parallel=\nhat\cdot \ve \nabla\tr(\ma S^2)\tauK^2\ellf$ and $X_\perp=\phat\cdot \ve \nabla\tr(\ma S^2)\tauK^2\ellf$, i.e. showing theory (solid lines), results for uniform initial positions (hollow markers) and by sequential update on time scale $\Tu=\Tp$ (filled markers).
	({\bf f}) Average squared strain following the strategy that optimizes the time average Eq.~(\ref{eq:TimeAverageStrain}) based on Eq.~(\ref{eq:TrSSqrConditionalGeneral}) for general values of initial strain and strain gradients against different time scales $T$.
	The solid line shows results for Gaussian distributed initial flow components with prediction horizon $\Tp=T$.
	Hollow markers and filled markers with $\Tp=\Tu=T$ same as in panel ({\rm e}).
	Also shown are results with small $\Tp$ ($\Tp=0.01\tauf$, $\Tu=T$) and small $\Tu$ ($T_{\rm u}=0.01\tauf$, $T_{\rm p}=T$).
	Parameters $\vsmax\tauf/\ellf=1$, $\omegasmax\tauf=5$, $\lambda=1$, and $\ku=0.1$.
	}
\end{figure}

Averaging Eq.~(\ref{eq:TrSSqrConditionalGeneral}) conditional on $X_\parallel$ gives
\begin{align}
	\label{eq:TrSSqrConditionalX}
	 & \langle\trSSt|X_\parallel\rangle =
	\frac{1}{2} + e^{-2t/\tauf-R^2}\bigg(
	\frac{\vs}{96\ellf\omegasq}\sin(\omegasq t)(96-144R^2+64R^4-12R^6+R^8)X_\parallel
	\\&
	+\frac{1}{192\sqrt{3}}\Big[96+96R^2-168R^4+80R^6-13R^8+R^{10}-\bigg[\frac{\ellf\omegasq}{2\vs}\bigg]^2R^4(288-288R^2+96R^4-16R^6+R^8)\Big] \Big(|X_\parallel|-\frac{\sqrt{3}}{2}\Big)
	\bigg)\,,
	\nonumber
\end{align}
and averaging conditional on $X_\perp$ gives
\begin{align}
	\label{eq:TrSSqrConditionalY}
	 & \langle\trSSt|X_\perp\rangle =
	\frac{1}{2} + e^{-2t/\tauf-R^2}\bigg(
	\frac{\vs}{96\ellf\omegasq}[1-\cos(\omegasq t)](96-144R^2+64R^4-12R^6+R^8)X_\perp
	\\&
	+\frac{1}{192\sqrt{3}}\Big[96-192R^2+120R^4-16R^6+3R^8+\bigg[\frac{\ellf\omegasq}{2\vs}\bigg]^2R^4(288-288R^2+96R^4-16R^6+R^8)\Big] \Big(|X_\perp|-\frac{\sqrt{3}}{2}\Big)
	\bigg)\,.
	\nonumber
\end{align}
Both averages have a linear contribution $\propto X_i$, and one contribution $\propto |X_i|$ with its mean $\sqrt{3}/2$ subtracted.
The first contribution is more sensitive to the choice of $\vs$ and $\omegasq$, meaning that it gives the dominant contribution to the optimal strategy.
The corresponding optimal strategies for choosing $\vsopt$ and $\omegasopt$ for different values of the signal are shown as solid lines in Fig.~\ref{fig:GaussConditional}({\bf a}--{\bf d}).

The strategy based on $X_\parallel$ in Eq.~(\ref{eq:TrSSqrConditionalX}) is to swim only if $X_\parallel$ is negative, i.e. if the strain decreases in the direction of swimming.
For small $\Tp$, swimming occurs at maximal speed, same as the strategy for $\vsopt$ in Eq.~(\ref{eq:optimal_policy}).
To not overshoot the low-strain region ahead, the swimming velocity lies below the maximum when optimizing on larger time scales.
Angular swimming is essentially zero, unless the gradient of $\trSS$ is large and negative, where a slight angular swimming is preferred.
Due to symmetries, swimming with positive or negative angular velocities give the same performance (Fig.~\ref{fig:GaussConditional}({\bf c}) shows the case of positive angular swimming).

The strategy based on $X_\perp$ in Eq.~(\ref{eq:TrSSqrConditionalY}) is to always swim, and to rotate in the opposite direction of the sign of $X_\perp$.
This is the same as the strategy for $\omegas_{q,{\rm opt}}(t)$ in Eq.~(\ref{eq:optimal_policy}).
As found in Ref.~\cite{mousavi2024efficient}, this turns the swimmer towards the direction where strain decreases the most, akin to gradient descent with a delay.
The magnitude of the optimal rotational swimming lies somewhat below $\omegasmax$ for large $X_\perp$ to avoid overshooting the optimal orientation.
For small $|X_\perp|$, the optimal magnitude of $\vs$ is smaller than its maximal value.
In this limit, there is a competition between the two contributions in Eq.~(\ref{eq:TrSSqrConditionalY}), where the second contribution dominates when $X_\perp=0$.
The strategy optimizing the second contribution turns out to be to swim if $|X_\perp|>\sqrt{3}/2$ and not swim otherwise, explaining why $\vsopt$ approaches zero for small $X_\perp$.

In the limit of small prediction times, the optimal strategies simplifies to
\begin{align}
	\label{eq:optimal_policy_X}
	\vsopt(X_\parallel) & =\left\{
	\begin{array}{ll}
		\vsmax & \mbox{if }X_\parallel<0\cr
		0      & \mbox{otherwise}
	\end{array}
	\right.\hspace{0.5cm}\mbox{and}\hspace{0.5cm}\omegasopt(X_\parallel)=0\,.
	\\
	\vsopt(X_\perp)     & =
	\left\{
	\begin{array}{ll}
		\frac{\sqrt{3}}{4}\frac{|X_\perp|}{\sqrt{3}/2-|X_\perp|}\ellf\omegasmax & \mbox{if }0\le\frac{\sqrt{3}}{4}\frac{|X_\perp|}{\sqrt{3}/2-|X_\perp|}\ellf\omegasmax<\vsmax\cr
		\vsmax                                                                  & \mbox{otherwise}
	\end{array}
	\right.\,,
	\hspace{0.5cm}\mbox{and}\hspace{0.5cm}
	\omegasopt(X_\perp)=\left\{
	\begin{array}{ll}
		\omegasmax  & \mbox{if }X_\perp<0\cr
		-\omegasmax & \mbox{otherwise}
	\end{array}
	\right.\,.
	\label{eq:optimal_policy_Y}
\end{align}
These are shown as thick black lines in Fig.~\ref{fig:GaussConditional}({\bf a}--{\bf d}).
They agree well with the optimal strategy for $\Tp=0.1\tauf$ (red curves).
The small adjustments from these strategies observed for larger $\Tp$ in Fig.~\ref{fig:GaussConditional}({\bf a}--{\bf d}) improves the predicted performance, but only by a few percent.

Fig.~\ref{fig:GaussConditional}({\bf e}) shows the performance of the optimal strategies for the signals $X_\parallel$ and $X_\perp$.
Solid lines are obtained by averaging Eqs.~(\ref{eq:TrSSqrConditionalX}), and (\ref{eq:TrSSqrConditionalY}) over time and over Gaussian distributed flow components. Hollow markers show corresponding simulation results.
We observe good agreement.
Similar to Fig.~\ref{fig:GaussConditionalStrain}({\bf b}), the filled markers show results for the scenario in which the swimmer sequentially updates its strategy with update time scale $\Tu=\Tp$. Preferential sampling improves the performance for the signal $X_\parallel$ and even more so for $X_\perp$, resulting in a lower average squared strain if $\Tp<\tauK$.
Fig~\ref{fig:GaussConditional}({\bf f}) shows the corresponding results for the optimal strategy obtained by averaging Eq.~(\ref{eq:TrSSqrConditionalGeneral}) where all components of $\ma S_0$ and $\ve\nabla\ma S_0$ in the $\nhat_0$, $\phat_0$ basis are used as signals.
When averaged over Gaussian distributed initial conditions, the performance is slightly better than for $X_\parallel$ or $X_\perp$ individually (solid lines, hollow markers).
Fig~\ref{fig:GaussConditional}({\bf f}) additionally shows results for swimmers updating their strategy with a short-sighted prediction horizon, $\Tp=0.01\tauf$, with variable update times $\Tu$.
As expected, the strategy performs well for small $\Tu$, while for larger $\Tu$, it is worse than strategies based on longer prediction horizons.
Using a short update time $\Tu=0.01\tauf$ with a large prediction time, on the other hand, significantly improves the performance, with a high efficiency around $\Tp=\tauf$.

Finally, since swimmers in nature do not have the ability to make perfect measurements, we have confirmed that the strategy in Fig~\ref{fig:GaussConditional}({\bf f}) is robust to measurement noise, even for a large prediction time $\Tp=\tauf$ with a small update time $\Tu=0.01\tauf$, see Appendix~\ref{app:noise} for details.
We find that the strategy remains effective even in the presence of measurement noise, with its efficiency declining only when the noise magnitude becomes comparable to the signal.

\section{Discussion}
\label{sec:discussion}

\subsection{Strategies based on squared strain}
\begin{figure}
	\includegraphics[width=0.4\textwidth]{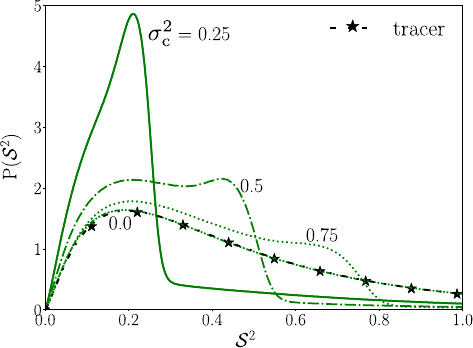}
	\caption{\label{fig:RL_consistency}
		Distribution of $\trSS$ in the statistical model for swimmers following Eq.~(\ref{eq:optimal_policy_TrSSqr}) with different thresholds $\sigma_{\rm c}^2=0$, $0.25$, $0.5$, and $0.75$ (green curves).
The distribution for tracer particles is shown as black dashed, $\star$.
		Same parameters as in Fig.~\ref{fig:RL_performance}({\bf a}) and $\PhiK=30$.
}
\end{figure}
The analysis in Section~\ref{sec:ApplicationTrSSqr} shows that the strategy in Eq.~(\ref{eq:optimal_policy_TrSSqr}) is optimal for any prediction horizon.
This, however, contradicts that the results from reinforcement learning are better, Fig.~\ref{fig:RL_performance}({\bf a}).
The strategy in Eq.~(\ref{eq:optimal_policy_TrSSqr}) is to swim if the initial strain lies above the strain $\sigma_{\rm c}^2=1/2$ of tracer particles. Thus, if the initial strain is above the expected strain, it is beneficial to swim to decorrelate from the initial strain as quickly as possible.
The strategy in Table~\ref{tab:strategy} is essentially the same, but with a lower threshold $\sigma_{\rm c}^2$ for swimming.
The difference can be understood by examination of the distribution of squared strain $\trSS$, shown in Fig.~\ref{fig:RL_consistency}.
It shows the distribution for cruising swimmers in the statistical model following Eq.~(\ref{eq:optimal_policy_TrSSqr}) with $\PhiK=30$ and different thresholds $\sigma_{\rm c}^2$ (green).
When following this strategy with $\sigma_{\rm c}^2=1/2$, one obtains a distribution (green dash-dotted line) that is significantly biased towards lower strain compared to the distribution of tracer particles (black dashed, $\star$).
A better strategy is therefore to instead swim above a threshold given by the average strain obtained from this preferentially sampled distribution.
However, since this new strategy is expected to further bias the sampled distribution towards lower strain, it could further reduce the value of the optimal threshold.
This is a self-consistency problem, by reducing the threshold, the preferentially sampled average is reduced, which in turn suggests a lower threshold.
This process can be continued until the preferentially sampled average no longer reduces, which happens when the preferentially sampled strain is equal to the threshold, $\langle\trSSt\rangle=\sigma^2_{\rm c}$.
Reinforcement learning successfully solves this self-consistency problem by finding a solution on the form in Eq.~(\ref{eq:optimal_policy_TrSSqr}) with an optimal threshold $\sigma^2_{\rm c}$ adjusted for preferential sampling.
The threshold $\sigma^2_{\rm c}=0.25$ lies close to the optimal threshold in Table.~\ref{tab:strategy}, and is plotted as solid green.
The distribution is strongly skewed, with a high peak below the threshold, and low probability for strain larger than the threshold.
Decreasing the threshold further is counterproductive, leading to a distribution with larger average $\trSS$.
As the threshold approaches zero, the distribution lies very close to that of tracer particles.

The bias in the distribution towards lower strain discussed above is the same mechanism explaining why the solid markers in Fig.~\ref{fig:GaussConditionalStrain}({\bf b}) for small $\Tp$ takes smaller values than the hollow markers.
We remark that the results in Fig.~\ref{fig:GaussConditionalStrain}({\bf b}) were evaluated using small $\ku$ for $\vsmax\tauf/\ellf=1$. Replacing the time scale $\tauf\to\tauK$ in this relation gives $\PhiK\approx 6$, allowing for comparison to Fig.~\ref{fig:RL_performance}({\bf a}), where $\PhiK=6$ gives $\langle\trSSt\rangle\sim0.42$, of the same order as the value in Fig.~\ref{fig:RL_performance}({\bf a}).

\subsection{Strategies based on squared strain and its derivatives}

The analysis in Section~\ref{sec:ApplicationGradient}  explains that the gradients of squared strain, $X_\parallel$ and $X_\perp$, are of prime importance for avoiding high-strain regions, and the learning works best when the prediction time is of the order of $\tauf$. This follows from Fig.~\ref{fig:GaussConditional}({\bf e}) which shows
that  the predicted average strain has a minimum at $\Tp=\tauf$ [see also Fig.~\ref{fig:GaussConditionalStrain}({\bf b})].
This shows that the strategies efficiently exploit flow correlations on this time scale. For larger times, the flow decorrelates from the initial condition, meaning that prediction is no longer possible.
The strategies based on $X_\parallel$ and $X_\perp$ have similar performance, but $X_\perp$ is slightly better if the strategy is sequentially updated [solid markers in Fig.~\ref{fig:GaussConditional}({\bf e})].
These two signals also give better performance than the optimal strategy based on strain in Fig.~\ref{fig:GaussConditionalStrain}({\bf b}), showing that strain gradients contain more relevant information for navigation.
Comparison to Fig~\ref{fig:GaussConditional}({\bf f}) shows that including all signals improves the predicted strain (solid lines).
However, when sampled along swimming trajectories with sequentially updated strategy, the performance is on the same level, or slightly worse than the strategy based on $X_\perp$ only.
The reason why preferential sampling gives very good performance for that strategy is that it tends to align the swimmer such that strain is decreasing along the swimmer direction at the times where the strategy is updated.
Thus, when preferential sampling is included, it is advantageous to not only optimize $\trSS$, but also optimize the configuration that is obtained in the next update, allowing for even lower strain sampled in the long run.
It is hard to find such optimal strategies in general, because when preferential sampling is included, the strategy based on Gaussian flow components in Eq.~(\ref{eq:TrSSqrConditionalGeneral}) is no longer exact.
Reinforcement learning takes this into account.

Fig.~\ref{fig:GaussConditional}({\bf f}) shows that a short update time $\Tu\ll\tauf$, leads to an optimum in the performance at a prediction horizon $\Tp\sim\tauf$ (blue,$\vartriangleleft$). Larger values of $\Tp$ leads to worse performance.
This behavior is similar to that observed in Ref.~\cite{monthiller2022surfing}, where the navigation performance was evaluated numerically based on a small $\Tu$ and with estimates based on different time horizons. Also there, an optimum was found around the flow time scale.
In these cases, the prediction made on the time scale $\Tp$ assumes that the action is kept constant. This neglects the contribution due to update at each $\Tu$, resulting in a poor prediction and hence strategy at very large $\Tp$.
In reinforcement learning by contrast, the update of the swimming behavior is included in the prediction, meaning that there is no harm to put a long time horizon, as long as the algorithm converges.
We therefore expect that the performance keep increasing, or reach a plateau at prediction horizons larger than the time scale of the flow.

Using the expansion in Eq.~(\ref{eq:TrSSqrConditionalGeneral}), it is straightforward to derive the optimal strategy for additional signal combinations by integrating Eq.~(\ref{eq:TrSSqrConditionalGeneral}).
For example, we find that one can use signals that are not correlated to the instantaneous strain, such as $\sum_{i,j,k}[\partial_k S_{0,ij}]^2$, to reduce sampling of high-strain regions.
Such strategies instead rely on the spatial correlations with the strain, similar to active alignment of swimmers with the flow velocity in turbulence~\cite{borgnino2019alignment}.

\section{Conclusions}
\label{sec:conclusions}

We developed a new analytical approach to find efficient strategies for microswimmers to avoid high-strain regions in turbulent flows.
Starting from a Gaussian distribution of flow components, we analytically derived the true optimal strategy to minimize the average strain along trajectories of swimmers on an arbitrary time horizon.
Using first- and second-order flow gradients that are attainable of the local frame of a swimmer, we found the optimal strategy for a number of signal combinations.
The theory (\ref{eq:TrSSqrConditionalGeneral}) shows that if only first-order flow gradients are available, the squared strain $\sigma^2$ is the best signal.
The optimal strategy is to swim in the instantaneous direction when $\sigma^2$ lies above a threshold $\sigma_{\rm c}^2$.
This threshold becomes $\sigma_{\rm c}^2=1/2$ in the limit where the theory is exact. For swimmers with sequential updates of the strategy, the threshold can be found through reinforcement learning, or by using a self-consistency approach by refining $\sigma_{\rm c}^2$ until it agrees with the average $\langle\trSSt\rangle$ obtained when following the strategy.
If in addition, second-order gradients are available, the theory predicts that the best signals are spatial derivatives of squared strain, projected on the directions in the frame of the swimmer.
In this case, the strain signal $\sigma^2$ does not matter much.
The resulting strategies are similar to those obtained by earlier methods using short-time expansions and reinforcement learning, but there are differences.
For example, Fig.~\ref{fig:GaussConditional}({\bf f}) shows that the short-time strategy (green,$\star$) does not perform as well as the strategy with a finite prediction horizon $\Tp$ (magenta,$\vartriangle$) if the update time $\Tu$ is not kept small.
Another important advantage of the new method is that it is parameterized in terms of the signals, swimming abilities and prediction time, Eq.~(\ref{eq:TrSSqrConditionalGeneral}).
This allows to quickly identify the most important signals and swimming abilities for different purposes.

We optimized the analytical optimal strategy on different time horizons $\Tp$.
The results show that it is possible to exploit flow correlations to reduce the strain levels up to times of the order of the flow time scale $\tauf$ in the statistical model.
Moreover, if the strategy is used to update the swimming behavior at regular time intervals $\Tu$, the resulting preferential sampling improves the performance.
It is an open question how one can include this active preferential sampling in the prediction. But at least for small Ku, we expect that  approaches similar to those reviewed in Ref.~\cite{gustavsson2016statistical} can resolve this problem.

The found reinforcement-learning strategy offers a perspective regarding the behavioural adaptation  of microorganisms in the ocean to different flow conditions.
It is shown in Ref.~\cite{gilbert2005turbulence} that the threshold triggering the response differs in calm and turbulent environments, the swimmers tend to be less sensitive to hydrodynamic signals in more intensive turbulence. The authors argued that this behaviour reduces energy consumption~\cite{gilbert2005turbulence}.
Our predicted strain [Eq.~(\ref{eq:TrSSqrConditionalTrSSqr})] provides an alternative explanation. Given only the magnitude of strain rate, the optimal strategy to avoid high strain regions is to swim whenever the strain rate is above the average sampled by the swimmers [Eq.~(\ref{eq:optimal_policy_TrSSqr})].
As explained above, preferential sampling leads to a self-consistency solution where the threshold agrees with the average strain.
The final level, however, depends on the turbulence intensity and is larger in more intensive turbulence.
This may explain the reduced sensitivity to hydrodynamic disturbance in the presence of intensive turbulence.
Second, we find that the optimal time-horizon is of the order of the characteristic time scale of turbulent fluctuations. This time scale can be very different in calm and highly
turbulent environments, and our results offer a possible reason for the observed difference in behaviour in these two cases.

We studied the single task of avoiding high-strain regions using flow velocity gradients as signals.
This is a simplified problem, and it is expected that swimmers in nature and in applications may have access to additional signals and optimize additional goals, once the high-strain regions are avoided.
A largely open question is how strain, or other signals, such as chemical concentrations, light, pressure, or slip velocity due to settling, can be used to solve competing goals.
Recently, such problems have been addressed for microswimmers in turbulent environments using reinforcement learning~\cite{calascibetta2023taming,xu2023long} and optimal control~\cite{piro2023energetic}. It would be interesting to use the analytical method introduced here to address such problems.

\begin{acknowledgments}
	We acknowledge support from Vetenskapsr\aa{}det,  grant nos. 2018-03974, 2023-03617 (JQ and KG), and 2021-4452 (BM). KG, BM and JQ acknowledge
	support from the Knut and Alice Wallenberg Foundation, grant no. 2019.0079.
	JQ and LZ were supported by the National Natural Science Foundation of China (grant nos. 92252104 and 12388101).
	Statistical-model simulations were performed on resources provided by the Swedish National Infrastructure for Computing (SNIC), partially funded by the Swedish Research Council through grant agreement no. 2018-05973.

\end{acknowledgments}

\appendix

\section{Flow models}
\label{app:flow}

\subsubsection{DNS of turbulence}
\label{sec:DNS}
Incompressible homogeneous isotropic turbulence is simulated using the same code as in Refs.~\cite{qiu2022active,mousavi2024efficient}. A pseudo-spectral method is used to solve the Navier-Stokes equations
\begin{align}
	\frac{{\partial {\ve u}}}{{\partial t}} + \ve u \cdot \ve\nabla\ve u =  - \frac{1}{{{\rho_{\rm f}}}}\ve\nabla {p_{\rm f}} + \nu {\nabla ^2}\ve u + \ve f\,,\mbox{ with }  \ve\nabla  \cdot \ve u = 0\,,
	\label{eq:NS}
\end{align}
where $p_{\rm f}$, $\rho_{\rm f}$ and $\nu$ are the pressure, density and kinematic viscosity of the fluid, respectively. An external force $\ve f$ is applied to balance the energy dissipation of turbulence using the method in Ref~\cite{machiels1997predictability}. Periodic boundary conditions are applied to all boundaries of a cubic domain.

The turbulence intensity is quantified by the Taylor Reynolds number, ${\rm Re}_\lambda = \sqrt{5/3}\langle\ve u^2\rangle\tauK^2/\eta^2$, with Kolmogorov time scale $\tauK$ and length scale $\eta$. The simulations in the present study have Re$_\lambda\approx 60$. We use a $96^3$ mesh for the domain with a size of $(2\pi)^3$ to resolve the flow field. The smallest resolved scale is $\eta/1.78$, which ensures that the finest turbulent motion is resolved. A statistical model flow field with exponential energy spectrum is used for the initial field, and Eqs. (\ref{eq:NS}) are integrated by a second-order Adams-Bashforth scheme with a time step of approximately $0.01\tauK$. After the turbulence is fully developed, swimmers are initialized at random positions, whose trajectories are calculated by interpolating fluid velocity and its gradients at the position of swimmers using second-order Lagrangian interpolation.

In the simulations needed for reinforcement learning, we use the entire time-dependent flow to train in a temporally evolving turbulent environment.
This is in contrast to our earlier studies~\cite{qiu2022navigation,qiu2022active}, where frozen flow snapshots from DNS were used during learning.

The short-time optimal strategies were further tested using DNS data of forced homogeneous isotropic turbulence with ${\rm Re}_\lambda\approx 418$ on a $1024^3$ grid, obtained from the Johns Hopkins University turbulence database~\cite{JohnsHopkins,JohnsHopkins2}, using the same setup as in Ref.~\cite{mousavi2024efficient}.
We integrated the swimmer dynamics using a second-order Adams-Bashforth method with a of approximately $0.005\tauK$. The flow at the particle positions was retrieved at regular time intervals and linearly interpolated in between.

\subsubsection{Statistical model}
We use a statistical model for the turbulent velocity fluctuations, writing $\ve u(\ve x,t)=\ve\nabla\cross\ve\Psi(\ve x,t)$. The vector potential~$\ve \Psi$ has zero mean and a homogeneous, isotropic correlation function on the form~\cite{gustavsson2016statistical,bec2024statistical}
\begin{align}
	\langle\Psi_i(\ve x,t)\Psi_j(\ve x',t')\rangle=\frac{1}{d(d-1)}\delta_{ij}\uf^2\ellf^2e^{-|\ve x-\ve x'|/(2\ellf^2)-|t-t'|/\tauf}\,.
	\label{eq:SMcorr}
\end{align}
Here $d=2,3$ denotes the spatial dimension. In two dimensions, the components $\Psi_1$ and $\Psi_2$ are defined to vanish.
The fluid-velocity field has a single length scale, $\ellf$, the  time scale $\tauf$, and the root-mean squared velocity $\uf=\sqrt{\langle\ve u^2\rangle}$.
The model is characterised by  the non-dimensional Kubo number, $\ku=\uf\tauf/\ellf$~\cite{wilkinson2007unmixing}.
For large values of $\ku$, model results agree well with DNS results for inertial particles~\cite{bec2024statistical} and microswimmers~\cite{gustavsson2016preferential,borgnino2022alignment,mousavi2024efficient} in turbulence.
When $\ku$ is small, relative time scales are different in the model compared to DNS, but the mechanisms underlying the particle dynamics is often the same~\cite{gustavsson2016statistical}.

For the model simulations discussed in Section~\ref{sec:analytical}, we consider a two-dimensional flow with $\ku=0.1$.
We use Gaussian distributed $\ve\Psi$ obtained from a Fourier series with random time-dependent coefficients~\cite{gustavsson2016statistical}.
The resulting flow velocities are Gaussian distributed.

In the reinforcement-learning simulations discussed in Section~\ref{sec:RL}, we use a three-dimensional flow with $\ku=10$.
To speed up simulations, we obtain the time-dependent flow
$\ve\Psi(\ve x,t)=\sum_{m=1}^mC_m(t)\ve\Psi_m(\ve x)$
by a superposition of $M=10$ pre-calculated flow snapshots $\ve\Psi_m(\ve x)$, with $m=1,\dots,M$ and independent Gaussian distributed coefficients $C_m$ following an Ornstein-Uhlenbeck process~\cite{mousavi2024efficient}.
For this case, flow components and flow gradients have non-Gaussian tails for finite $M$~\cite{meibohm2024caustic}.

To compare the results from the statistical model with large $\ku$ to DNS, we match the covariances of all components of fluid gradients, $\langle(\partial_ju_i)(\partial_lu_k)\rangle$, and second-order gradients, $\langle(\partial_j\partial_ku_i)(\partial_m\partial_nu_l)\rangle$, both being local quantities (in contrast to the flow velocity).
Symmetries constrain the forms of these covariances.
For homogeneous, isotropic, incompressible velocity fields, they can be written proportional to $\langle \tr(\ma S^2)\rangle$ and $\langle\sum_{p,q,r}(\partial_{q}\partial_ru_{p})^2\rangle$, respectively.
Define
\begin{align}
\tauK&=\frac{1}{2\langle\tr(\ma S^2)\rangle^{1/2}}\hspace{0.5cm}\mbox{and}\hspace{0.5cm}\eta=\frac{C}{\tauK\langle\sum_{p,q,r}(\partial_{q}\partial_ru_{p})^2\rangle^{1/2}}
\end{align}
where $C$ is a dimensionless proportionality factor.
For the DNS of homogeneous isotropic turbulence described above, $C$ is independent of the Reynolds number up to numerical precision; $C=0.28$ for ${\rm Re}_\lambda=60$ and $C=0.26$ for ${\rm Re}_\lambda=418$.
For the statistical model $\langle\tr(\ma S^2)\rangle=5/2\uf^2/\ellf^2$ and $\langle\sum_{p,q,r}(\partial_{q}\partial_ru_{p})^2\rangle=35/2\uf^2/\ellf^4$, giving $\tauK=\ellf/(\sqrt{5}\uf)$ and $\eta=\sqrt{2/7}C\ellf$.
When matching the statistical model to DNS we use $\eta=0.15\ellf$.

\section{Reinforcement learning}
\label{app:RL}
We use the one-step Q-learning algorithm because it allows to directly read off and interpret the optimal strategy.
This algorithm is based upon a table $Q(s,a)$ which, upon convergence, gives the expected discounted future reward for each state $s$ and action $a$.
The optimal policy is then to, for each state, choose the action with the largest $Q(s,a)$.
To converge to the optimal policy, $Q$ is updated by
\begin{equation}
	Q(s,a) \leftarrow Q(s,a) + \alpha [r + \gamma \max_{a'} Q(s',a') - Q(s,a)],
\end{equation}
each time the state changes during the training.
Here $a$ was the action taken in the previous state $s$, $s'$ denotes the new state, $a'$ is the action with maximal $Q$ in the new state, and $r$ denotes the reward.
Moreover, $\alpha$ is the learning rate, and $\gamma$ is the discount factor that determines the number of state changes upon which the policy is optimized.

For the jumping swimmer in the DNS, we update the action at constant time intervals $\Tu=0.044\tauK$, regardless of whether the state has changed.
This time is slightly longer than the duration of a single jump, allowing the copepod to swim by making successive jumps.
We choose $\gamma=0.99$, leading to a time horizon of optimization of the order $\Tu/(1-\gamma)\sim 4\tauK$, much larger than the update time, and somewhat larger than the smallest time scale of the flow.
For the cruising swimmer in the statistical model, we update the action each time a new state is encountered. This happens at irregular time intervals.
However, we choose a larger $\gamma=0.999$, ensuring that simulations are optimized on a time scale much larger than $\tauK$.

The training is carried out with an $\epsilon$-greedy policy, which implies taking random actions with probability $\epsilon$ during training to explore the state space and prevent convergence to suboptimal policies. We decrease $\epsilon$ linearly with the current episode number $E$, $\epsilon = \epsilon_0 \frac{E_{\epsilon} - E}{E_{\epsilon}}$, until it reaches zero at episode $E_{\epsilon}$.
We also reduce the learning rate with the episode according to $\alpha=\alpha_0/(1+E/E_\alpha)$ with episode scale $E_\alpha$.
The parameters used in the different training cases are stated in Table.~\ref{tab:training_params}.

\begin{table}
	\begin{tabular}{llllll}
		\hline\hline
		Parameter                 & Symbol         & Cruising    & Jumping\cr\hline
		Initial learning rate     & $\alpha_0$     & $0.2$       & $[0.008, 0.12]$\cr
		Learning rate decay scale & $E_\alpha$     & 1000        & 100\cr
		Discount factor           & $\gamma$       & $0.999$     & $0.99$\cr
		Initial exploration rate  & $\epsilon_0$   & $0.2$       & $0.05$\cr
		Exploration episode count & $E_{\epsilon}$ & 2000        & 250 \cr
		Episode duration          &                & $6700\tauK$ & $180\tauK$ \cr
		Number of episodes        &                & 3000        & 400\cr\hline\hline
	\end{tabular}
	\caption{\label{tab:training_params}Training parameters for cruising and jumping swimmers. Here $\tauK$ is the Kolmogorov time.
	}

\end{table}

\section{Evaluation of the short-time strategies for cruising swimmers}
\label{app:cruiser}
\begin{figure}
	\includegraphics[width=0.8\textwidth]{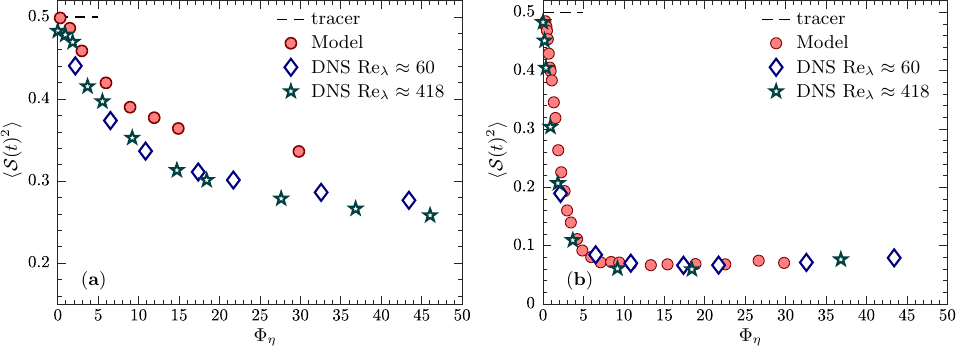}
	\caption{\label{fig:cruiser}
Average squared strain $\langle\trSSt\rangle$ for cruising swimmers with swimming speed $\PhiK$ following the strategies in ({\bf a}) Eq.~(\ref{eq:optimal_policy_TrSSqr}) and ({\bf b}) Eq.~(\ref{eq:optimal_policy}).
Model data (red,$\bullet$) is the same as in Fig.~\ref{fig:RL_performance}.
Hollow markers show results from DNS with ${\rm Re}_\lambda\approx 60$ (navy blue,$\Diamond$) and ${\rm Re}_\lambda\approx 418$ (dark teal,\raisebox{-0.7mm}{\FiveStarOpen}).
Parameters as in Fig.~\ref{fig:RL_performance}.
	}
\end{figure}

Figure~\ref{fig:cruiser} compares numerical simulations of the short-time strategies in Eqs.~(\ref{eq:optimal_policy_TrSSqr}) and~(\ref{eq:optimal_policy}) using flows obtained  from both statistical model simulations and DNS for two values of ${\rm Re}_\lambda$, as described in Appendix~\ref{app:flow}.

\section{Sensitivity of optimal strategy to measurement noise}
\label{app:noise}
\begin{figure}
	\includegraphics[width=0.35\textwidth]{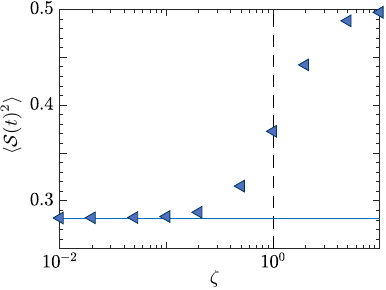}
	\caption{\label{fig:noise}
Average squared strain $\langle\trSSt\rangle$ against magnitude of measurement noise $\zeta$ (see text).
Parameters as in Fig.~\ref{fig:GaussConditional}({\bf f}) with $\Tp=\tauf$ and $\Tu=0.01\tauf$.
Horizontal line shows the value from Fig~\ref{fig:GaussConditional}({\bf f}) where the noise is absent ($\zeta=0$).
	}
\end{figure}
To confirm that the optimal strain-avoidance strategy based on all signals shown in Fig. 4(f) remains efficient in the presence of measurement noise, we performed simulations where Gaussian noise was added to each signal at every strategy update.
Each noise had standard deviation equal to $\zeta$ times the standard deviation of the corresponding signal.
The results for the prediction time $\Tp=\tauf$ with a small update time $\Tu=0.01\tauf$ are shown in Figure~\ref{fig:noise}.
We find that performance remains unaffected for $\zeta$ up to 0.1. Even for $\zeta=1$, where noise equals signal strength, the strategy remains efficient, though performance begins to decline. For $\zeta=10$, the signal is entirely masked, and the swimmer does not perform better than random sampling of strain.

\section{Alternative reinforcement-learning setups for navigation based on strain magnitude}
\label{app:DQN}
The reinforcement learning results in Section~\ref{sec:RLstrain} is based on Q-learning, which assumes discretized signals and actions.
To include continuous signals or actions other schemes must be employed.
Here we present results from training with a continuous strain magnitude $\sigma^2$ obtained using deep Q-learning (DQN)~\cite{mnih2015human} and deep deterministic policy gradients (DDPG)~\cite{lillicrap2015continuous}.

DQN is similar to Q-learning but uses a neural network instead of a look-up table. We use a single-layer perceptron with 16 nodes. The input is the value of $\sigma^2$ at the position of the swimmer, and the output layer has 30 nodes, each representing a discrete swimming speed linearly spaced from $0$ to $\vsmax$.
The chosen action is to swim at the speed associated with the node with the highest output value.

DDPG uses an actor-critic approach that allows continuous actions.
We use an actor network with two hidden layers of 18 and 32 neurons, and a critic network with two hidden layers of 10 and 32 neurons.
The actor network takes $\sigma^2$ as input and outputs a coefficient $c \in [0, 1]$ to set the swimming speed $v^{s} = c v_{\max}^{\text{(s)}}$ (deterministic policy).
The critic network then evaluates the Q-value for the state and resulting action from the actor, making it feasible to handle continuous actions by computing the Q-value for only the selected action.

\begin{figure}[h]
	\includegraphics[width=0.9\textwidth]{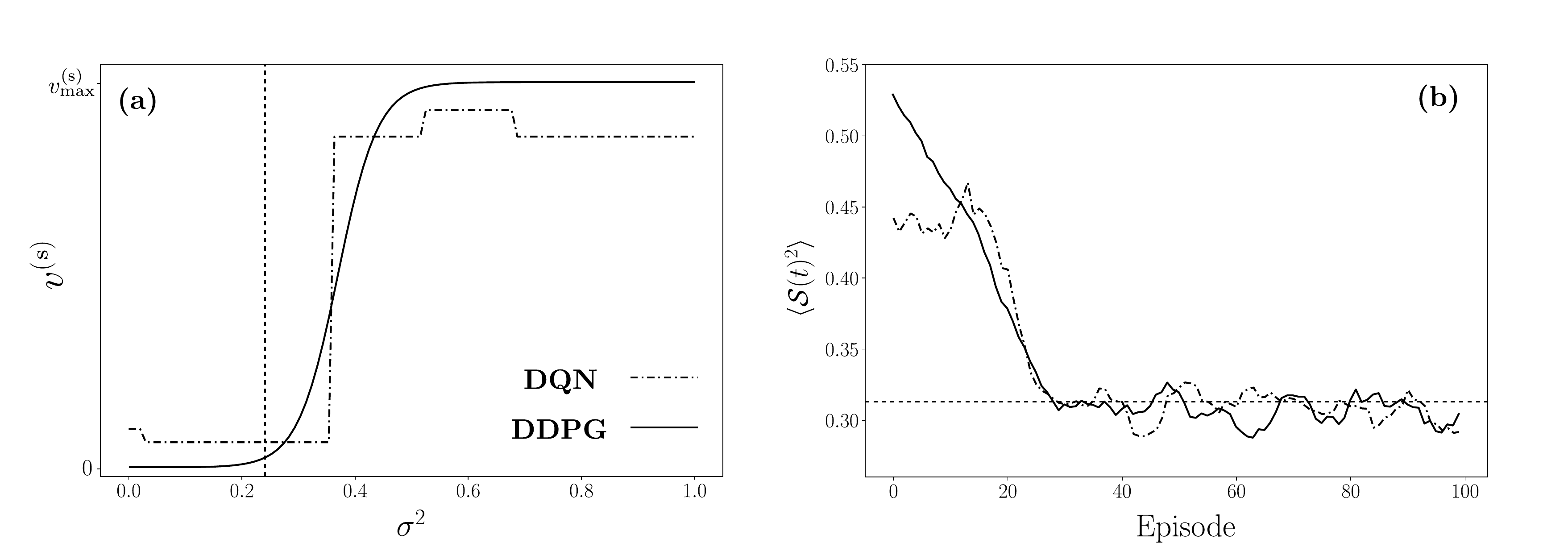}
	\caption{\label{fig:DQN}
		Results from deep reinforcement learning in the stochastic model: (\textbf{a}) control policy and (\textbf{b}) training performance for a continuous signal $\sigma^2$ with discretized action (DQN) and continuous actions (DDPG).
Dashed lines show (\textbf{a}) the threshold value from Q-learning in Table~\ref{tab:strategy}, and (\textbf{b}) the performance of the best Q-learning strategy.
Parameters $\PhiK=30$, $\Xi=0$ and $\lambda=2$.
	}
\end{figure}

Fig.~\ref{fig:DQN} shows results of training in the stochastic model using deep Q-learning (DQN)~\cite{mnih2015human} and deep deterministic policy gradients (DDPG)~\cite{lillicrap2015continuous}.
In both cases the optimal strategy is essentially to swim if the strain lies above a threshold, and not swim otherwise.
This is the same strategy and approximately the same threshold obtained by tabular Q-learning in Table~\ref{tab:strategy}. The performance for the three cases is of the same order, see Fig.~\ref{fig:DQN}(b).

The strategies in Sec.~\ref{sec:RLstrain} assumes that the rotational swimming is zero.
Table.~\ref{tab:strategy_sigma_rotational} shows the optimal control policy obtained through reinforcement learning including rotational swimming.
The results show a clear distinction between stop/swim at the same threshold value similar to Table.~\ref{tab:strategy}.
The chosen angular swimming lack clear structure and are nearly uniformly distributed.
The exception is at very high strain, where angular swimming is never chosen.
The average strain sampled by the swimmer is $\langle \trSS(t)^2\rangle = 0.31$, which is in the same order of the case without rotational control shown in Fig.~\ref{fig:RL_performance} (\textbf{a}).
We come to the same conclusion as the theoretical approach in Sec.~\ref{sec:ApplicationTrSSqr}: rotational swimming does not contribute much when navigating using $\sigma^2$ as the signal.

\begin{table}
	\begin{overpic}[width=0.85\textwidth]{{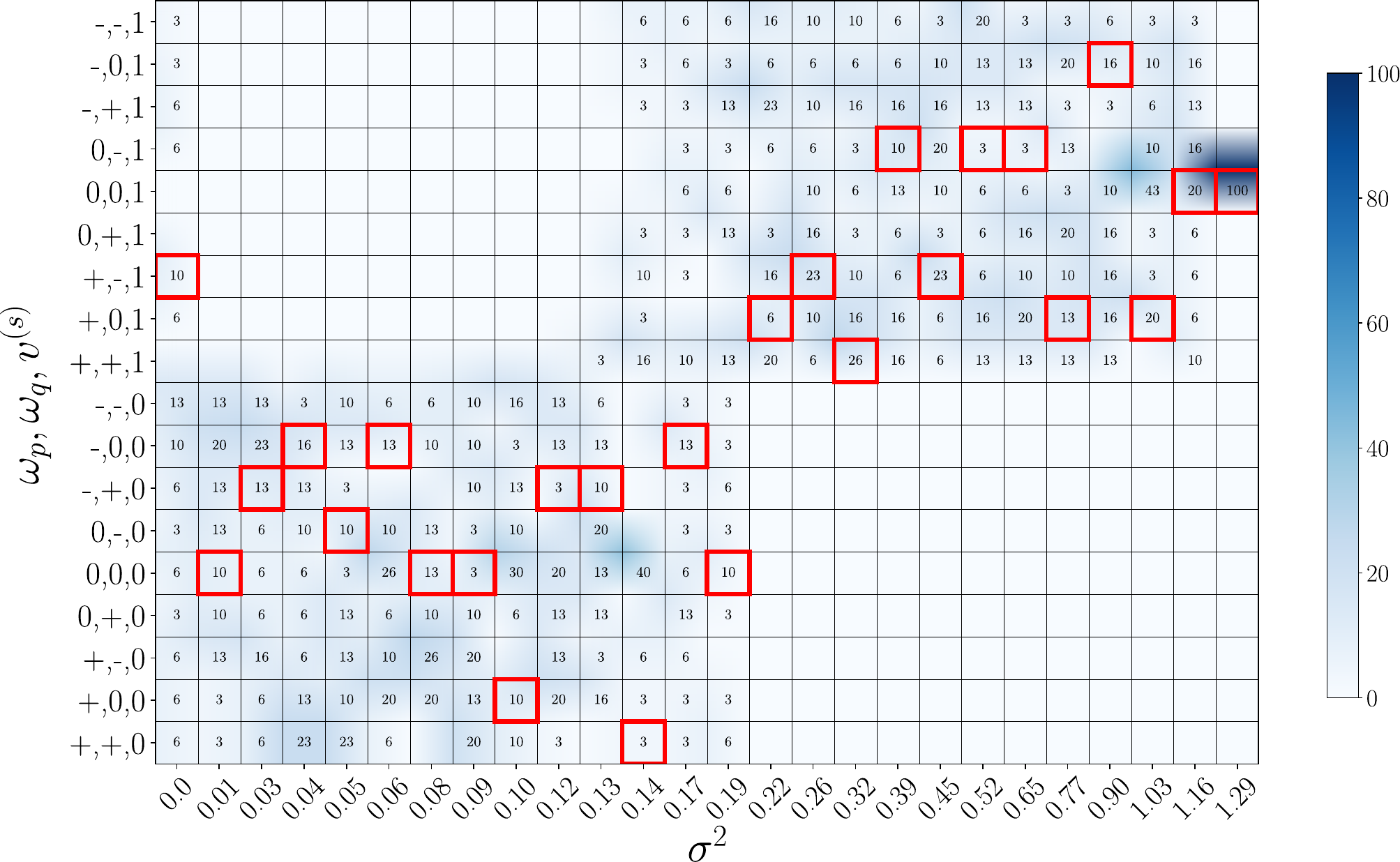}}
	\end{overpic}
	\caption{\label{tab:strategy_sigma_rotational}
		Results from Q-learning for cruising swimmers that sense strain magnitude, and exhibit both translational and rotational control.
Strain $\sigma^2$ is discretized as in Sec.~\ref{sec:RLstrain} and actions are discretized as in Sec.~\ref{sec:RLstrain_gradients}.
		Numbers give percentage of 30 trained strategies taking an action (row) in a state (column).
		Red frames show the best found strategy. Parameters: $\PhiK=30$ and $\Xi=6.25$.
	}
\end{table}

%\bibliography{strain}

\begin{thebibliography}{72}%
\makeatletter
\providecommand \@ifxundefined [1]{%
 \@ifx{#1\undefined}
}%
\providecommand \@ifnum [1]{%
 \ifnum #1\expandafter \@firstoftwo
 \else \expandafter \@secondoftwo
 \fi
}%
\providecommand \@ifx [1]{%
 \ifx #1\expandafter \@firstoftwo
 \else \expandafter \@secondoftwo
 \fi
}%
\providecommand \natexlab [1]{#1}%
\providecommand \enquote  [1]{``#1''}%
\providecommand \bibnamefont  [1]{#1}%
\providecommand \bibfnamefont [1]{#1}%
\providecommand \citenamefont [1]{#1}%
\providecommand \href@noop [0]{\@secondoftwo}%
\providecommand \href [0]{\begingroup \@sanitize@url \@href}%
\providecommand \@href[1]{\@@startlink{#1}\@@href}%
\providecommand \@@href[1]{\endgroup#1\@@endlink}%
\providecommand \@sanitize@url [0]{\catcode `\\12\catcode `\$12\catcode
  `\&12\catcode `\#12\catcode `\^12\catcode `\_12\catcode `\%12\relax}%
\providecommand \@@startlink[1]{}%
\providecommand \@@endlink[0]{}%
\providecommand \url  [0]{\begingroup\@sanitize@url \@url }%
\providecommand \@url [1]{\endgroup\@href {#1}{\urlprefix }}%
\providecommand \urlprefix  [0]{URL }%
\providecommand \Eprint [0]{\href }%
\providecommand \doibase [0]{https://doi.org/}%
\providecommand \selectlanguage [0]{\@gobble}%
\providecommand \bibinfo  [0]{\@secondoftwo}%
\providecommand \bibfield  [0]{\@secondoftwo}%
\providecommand \translation [1]{[#1]}%
\providecommand \BibitemOpen [0]{}%
\providecommand \bibitemStop [0]{}%
\providecommand \bibitemNoStop [0]{.\EOS\space}%
\providecommand \EOS [0]{\spacefactor3000\relax}%
\providecommand \BibitemShut  [1]{\csname bibitem#1\endcsname}%
\let\auto@bib@innerbib\@empty
%</preamble>
\bibitem [{\citenamefont {Jiang}\ and\ \citenamefont
  {Osborn}(2004)}]{jiang2004hydrodynamics}%
  \BibitemOpen
  \bibfield  {author} {\bibinfo {author} {\bibfnamefont {H.}~\bibnamefont
  {Jiang}}\ and\ \bibinfo {author} {\bibfnamefont {T.~R.}\ \bibnamefont
  {Osborn}},\ }\bibfield  {title} {\bibinfo {title} {Hydrodynamics of copepods:
  a review},\ }\href@noop {} {\bibfield  {journal} {\bibinfo  {journal}
  {Surveys in Geophysics}\ }\textbf {\bibinfo {volume} {25}},\ \bibinfo {pages}
  {339} (\bibinfo {year} {2004})}\BibitemShut {NoStop}%
\bibitem [{\citenamefont {Ki{\o}rboe}(2008)}]{kiorboe2008mechanistic}%
  \BibitemOpen
  \bibfield  {author} {\bibinfo {author} {\bibfnamefont {T.}~\bibnamefont
  {Ki{\o}rboe}},\ }\href@noop {} {\emph {\bibinfo {title} {A mechanistic
  approach to plankton ecology}}}\ (\bibinfo  {publisher} {Princeton University
  Press},\ \bibinfo {year} {2008})\BibitemShut {NoStop}%
\bibitem [{\citenamefont {Webster}\ and\ \citenamefont
  {Weissburg}(2009)}]{webster2009the}%
  \BibitemOpen
  \bibfield  {author} {\bibinfo {author} {\bibfnamefont {D.}~\bibnamefont
  {Webster}}\ and\ \bibinfo {author} {\bibfnamefont {M.}~\bibnamefont
  {Weissburg}},\ }\bibfield  {title} {\bibinfo {title} {The hydrodynamics of
  chemical cues among aquatic organisms},\ }\href@noop {} {\bibfield  {journal}
  {\bibinfo  {journal} {Annual Review of Fluid Mechanics}\ }\textbf {\bibinfo
  {volume} {41}},\ \bibinfo {pages} {73} (\bibinfo {year} {2009})}\BibitemShut
  {NoStop}%
\bibitem [{\citenamefont {Heuschele}\ and\ \citenamefont
  {Selander}(2014)}]{heuschele2014the}%
  \BibitemOpen
  \bibfield  {author} {\bibinfo {author} {\bibfnamefont {J.}~\bibnamefont
  {Heuschele}}\ and\ \bibinfo {author} {\bibfnamefont {E.}~\bibnamefont
  {Selander}},\ }\bibfield  {title} {\bibinfo {title} {The chemical ecology of
  copepods},\ }\href@noop {} {\bibfield  {journal} {\bibinfo  {journal}
  {Journal of Plankton Research}\ }\textbf {\bibinfo {volume} {36}},\ \bibinfo
  {pages} {895} (\bibinfo {year} {2014})}\BibitemShut {NoStop}%
\bibitem [{\citenamefont {Kessler}(1985)}]{kessler1985hydrodynamic}%
  \BibitemOpen
  \bibfield  {author} {\bibinfo {author} {\bibfnamefont {J.~O.}\ \bibnamefont
  {Kessler}},\ }\bibfield  {title} {\bibinfo {title} {Hydrodynamic focusing of
  motile algal cells},\ }\href@noop {} {\bibfield  {journal} {\bibinfo
  {journal} {Nature}\ }\textbf {\bibinfo {volume} {313}},\ \bibinfo {pages}
  {218} (\bibinfo {year} {1985})}\BibitemShut {NoStop}%
\bibitem [{\citenamefont {Guasto}\ \emph {et~al.}(2012)\citenamefont {Guasto},
  \citenamefont {Rusconi},\ and\ \citenamefont {Stocker}}]{guasto2012fluid}%
  \BibitemOpen
  \bibfield  {author} {\bibinfo {author} {\bibfnamefont {J.~S.}\ \bibnamefont
  {Guasto}}, \bibinfo {author} {\bibfnamefont {R.}~\bibnamefont {Rusconi}},\
  and\ \bibinfo {author} {\bibfnamefont {R.}~\bibnamefont {Stocker}},\
  }\bibfield  {title} {\bibinfo {title} {Fluid mechanics of planktonic
  microorganisms},\ }\href@noop {} {\bibfield  {journal} {\bibinfo  {journal}
  {Annual Review of Fluid Mechanics}\ }\textbf {\bibinfo {volume} {44}},\
  \bibinfo {pages} {373} (\bibinfo {year} {2012})}\BibitemShut {NoStop}%
\bibitem [{\citenamefont {Durham}\ \emph {et~al.}(2013)\citenamefont {Durham},
  \citenamefont {Climent}, \citenamefont {Barry}, \citenamefont {De~Lillo},
  \citenamefont {Boffetta}, \citenamefont {Cencini},\ and\ \citenamefont
  {Stocker}}]{durham2013turbulence}%
  \BibitemOpen
  \bibfield  {author} {\bibinfo {author} {\bibfnamefont {W.~M.}\ \bibnamefont
  {Durham}}, \bibinfo {author} {\bibfnamefont {E.}~\bibnamefont {Climent}},
  \bibinfo {author} {\bibfnamefont {M.}~\bibnamefont {Barry}}, \bibinfo
  {author} {\bibfnamefont {F.}~\bibnamefont {De~Lillo}}, \bibinfo {author}
  {\bibfnamefont {G.}~\bibnamefont {Boffetta}}, \bibinfo {author}
  {\bibfnamefont {M.}~\bibnamefont {Cencini}},\ and\ \bibinfo {author}
  {\bibfnamefont {R.}~\bibnamefont {Stocker}},\ }\bibfield  {title} {\bibinfo
  {title} {Turbulence drives microscale patches of motile phytoplankton},\
  }\href@noop {} {\bibfield  {journal} {\bibinfo  {journal} {Nature
  Communications}\ }\textbf {\bibinfo {volume} {4}},\ \bibinfo {pages} {1}
  (\bibinfo {year} {2013})}\BibitemShut {NoStop}%
\bibitem [{\citenamefont {Zhan}\ \emph {et~al.}(2014)\citenamefont {Zhan},
  \citenamefont {Sardina}, \citenamefont {Lushi},\ and\ \citenamefont
  {Brandt}}]{zhan2014accumulation}%
  \BibitemOpen
  \bibfield  {author} {\bibinfo {author} {\bibfnamefont {C.}~\bibnamefont
  {Zhan}}, \bibinfo {author} {\bibfnamefont {G.}~\bibnamefont {Sardina}},
  \bibinfo {author} {\bibfnamefont {E.}~\bibnamefont {Lushi}},\ and\ \bibinfo
  {author} {\bibfnamefont {L.}~\bibnamefont {Brandt}},\ }\bibfield  {title}
  {\bibinfo {title} {Accumulation of motile elongated micro-organisms in
  turbulence},\ }\href@noop {} {\bibfield  {journal} {\bibinfo  {journal}
  {Journal of Fluid Mechanics}\ }\textbf {\bibinfo {volume} {739}},\ \bibinfo
  {pages} {22} (\bibinfo {year} {2014})}\BibitemShut {NoStop}%
\bibitem [{\citenamefont {Gustavsson}\ \emph {et~al.}(2016)\citenamefont
  {Gustavsson}, \citenamefont {Berglund}, \citenamefont {Jonsson},\ and\
  \citenamefont {Mehlig}}]{gustavsson2016preferential}%
  \BibitemOpen
  \bibfield  {author} {\bibinfo {author} {\bibfnamefont {K.}~\bibnamefont
  {Gustavsson}}, \bibinfo {author} {\bibfnamefont {F.}~\bibnamefont
  {Berglund}}, \bibinfo {author} {\bibfnamefont {P.}~\bibnamefont {Jonsson}},\
  and\ \bibinfo {author} {\bibfnamefont {B.}~\bibnamefont {Mehlig}},\
  }\bibfield  {title} {\bibinfo {title} {Preferential sampling and small-scale
  clustering of gyrotactic microswimmers in turbulence},\ }\href@noop {}
  {\bibfield  {journal} {\bibinfo  {journal} {Physical Review Letters}\
  }\textbf {\bibinfo {volume} {116}},\ \bibinfo {pages} {108104} (\bibinfo
  {year} {2016})}\BibitemShut {NoStop}%
\bibitem [{\citenamefont {Lovecchio}\ \emph {et~al.}(2019)\citenamefont
  {Lovecchio}, \citenamefont {Climent}, \citenamefont {Stocker},\ and\
  \citenamefont {Durham}}]{lovecchio2019chain}%
  \BibitemOpen
  \bibfield  {author} {\bibinfo {author} {\bibfnamefont {S.}~\bibnamefont
  {Lovecchio}}, \bibinfo {author} {\bibfnamefont {E.}~\bibnamefont {Climent}},
  \bibinfo {author} {\bibfnamefont {R.}~\bibnamefont {Stocker}},\ and\ \bibinfo
  {author} {\bibfnamefont {W.~M.}\ \bibnamefont {Durham}},\ }\bibfield  {title}
  {\bibinfo {title} {Chain formation can enhance the vertical migration of
  phytoplankton through turbulence},\ }\href@noop {} {\bibfield  {journal}
  {\bibinfo  {journal} {Science Advances}\ }\textbf {\bibinfo {volume} {5}}
  (\bibinfo {year} {2019})}\BibitemShut {NoStop}%
\bibitem [{\citenamefont {Sengupta}\ \emph {et~al.}(2017)\citenamefont
  {Sengupta}, \citenamefont {Carrara},\ and\ \citenamefont
  {Stocker}}]{sengupta2017phytoplankton}%
  \BibitemOpen
  \bibfield  {author} {\bibinfo {author} {\bibfnamefont {A.}~\bibnamefont
  {Sengupta}}, \bibinfo {author} {\bibfnamefont {F.}~\bibnamefont {Carrara}},\
  and\ \bibinfo {author} {\bibfnamefont {R.}~\bibnamefont {Stocker}},\
  }\bibfield  {title} {\bibinfo {title} {Phytoplankton can actively diversify
  their migration strategy in response to turbulent cues},\ }\href@noop {}
  {\bibfield  {journal} {\bibinfo  {journal} {Nature}\ }\textbf {\bibinfo
  {volume} {543}},\ \bibinfo {pages} {555} (\bibinfo {year}
  {2017})}\BibitemShut {NoStop}%
\bibitem [{\citenamefont {Yen}\ \emph {et~al.}(1992)\citenamefont {Yen},
  \citenamefont {Lenz}, \citenamefont {Gassie},\ and\ \citenamefont
  {Hartline}}]{yen1992mechanoreception}%
  \BibitemOpen
  \bibfield  {author} {\bibinfo {author} {\bibfnamefont {J.}~\bibnamefont
  {Yen}}, \bibinfo {author} {\bibfnamefont {P.~H.}\ \bibnamefont {Lenz}},
  \bibinfo {author} {\bibfnamefont {D.~V.}\ \bibnamefont {Gassie}},\ and\
  \bibinfo {author} {\bibfnamefont {D.~K.}\ \bibnamefont {Hartline}},\
  }\bibfield  {title} {\bibinfo {title} {Mechanoreception in marine copepods:
  electrophysiological studies on the first antennae},\ }\href@noop {}
  {\bibfield  {journal} {\bibinfo  {journal} {Journal of Plankton Research}\
  }\textbf {\bibinfo {volume} {14}},\ \bibinfo {pages} {495} (\bibinfo {year}
  {1992})}\BibitemShut {NoStop}%
\bibitem [{\citenamefont {Ki{\o}rboe}\ \emph {et~al.}(1999)\citenamefont
  {Ki{\o}rboe}, \citenamefont {Saiz},\ and\ \citenamefont
  {Visser}}]{kioerboe1999hydrodynamic}%
  \BibitemOpen
  \bibfield  {author} {\bibinfo {author} {\bibfnamefont {T.}~\bibnamefont
  {Ki{\o}rboe}}, \bibinfo {author} {\bibfnamefont {E.}~\bibnamefont {Saiz}},\
  and\ \bibinfo {author} {\bibfnamefont {A.}~\bibnamefont {Visser}},\
  }\bibfield  {title} {\bibinfo {title} {Hydrodynamic signal perception in the
  copepod acartia tonsa},\ }\href@noop {} {\bibfield  {journal} {\bibinfo
  {journal} {Marine Ecology Progress Series}\ }\textbf {\bibinfo {volume}
  {179}},\ \bibinfo {pages} {97} (\bibinfo {year} {1999})}\BibitemShut
  {NoStop}%
\bibitem [{\citenamefont {Fields}\ \emph {et~al.}(2002)\citenamefont {Fields},
  \citenamefont {Shaeffer},\ and\ \citenamefont
  {Weissburg}}]{fields2002mechanical}%
  \BibitemOpen
  \bibfield  {author} {\bibinfo {author} {\bibfnamefont {D.~M.}\ \bibnamefont
  {Fields}}, \bibinfo {author} {\bibfnamefont {D.}~\bibnamefont {Shaeffer}},\
  and\ \bibinfo {author} {\bibfnamefont {M.~J.}\ \bibnamefont {Weissburg}},\
  }\bibfield  {title} {\bibinfo {title} {Mechanical and neural responses from
  the mechanosensory hairs on the antennule of gaussia princeps},\ }\href@noop
  {} {\bibfield  {journal} {\bibinfo  {journal} {Marine Ecology Progress
  Series}\ }\textbf {\bibinfo {volume} {227}},\ \bibinfo {pages} {173}
  (\bibinfo {year} {2002})}\BibitemShut {NoStop}%
\bibitem [{\citenamefont {P{\'{e}}cseli}\ and\ \citenamefont
  {Trulsen}(2016)}]{pecseli2016planktons}%
  \BibitemOpen
  \bibfield  {author} {\bibinfo {author} {\bibfnamefont {H.~L.}\ \bibnamefont
  {P{\'{e}}cseli}}\ and\ \bibinfo {author} {\bibfnamefont {J.~K.}\ \bibnamefont
  {Trulsen}},\ }\bibfield  {title} {\bibinfo {title} {Plankton's perception of
  signals in a turbulent environment},\ }\href@noop {} {\bibfield  {journal}
  {\bibinfo  {journal} {Advances in Physics: X}\ }\textbf {\bibinfo {volume}
  {1}},\ \bibinfo {pages} {20} (\bibinfo {year} {2016})}\BibitemShut {NoStop}%
\bibitem [{\citenamefont {Genin}\ \emph {et~al.}(2005)\citenamefont {Genin},
  \citenamefont {Jaffe}, \citenamefont {Reef}, \citenamefont {Richter},\ and\
  \citenamefont {Franks}}]{genin2005swimming}%
  \BibitemOpen
  \bibfield  {author} {\bibinfo {author} {\bibfnamefont {A.}~\bibnamefont
  {Genin}}, \bibinfo {author} {\bibfnamefont {J.~S.}\ \bibnamefont {Jaffe}},
  \bibinfo {author} {\bibfnamefont {R.}~\bibnamefont {Reef}}, \bibinfo {author}
  {\bibfnamefont {.}~\bibnamefont {Richter}},\ and\ \bibinfo {author}
  {\bibfnamefont {P.~J.}\ \bibnamefont {Franks}},\ }\bibfield  {title}
  {\bibinfo {title} {Swimming against the flow: a mechanism of zooplankton
  aggregation},\ }\href@noop {} {\bibfield  {journal} {\bibinfo  {journal}
  {Science}\ }\textbf {\bibinfo {volume} {308}},\ \bibinfo {pages} {860}
  (\bibinfo {year} {2005})}\BibitemShut {NoStop}%
\bibitem [{\citenamefont {Shang}\ \emph {et~al.}(2008)\citenamefont {Shang},
  \citenamefont {Wang},\ and\ \citenamefont {Li}}]{shang2008resisting}%
  \BibitemOpen
  \bibfield  {author} {\bibinfo {author} {\bibfnamefont {X.}~\bibnamefont
  {Shang}}, \bibinfo {author} {\bibfnamefont {G.}~\bibnamefont {Wang}},\ and\
  \bibinfo {author} {\bibfnamefont {S.}~\bibnamefont {Li}},\ }\bibfield
  {title} {\bibinfo {title} {Resisting flow - laboratory study of rheotaxis of
  the estuarine copepod pseudodiaptomus annandalei},\ }\href@noop {} {\bibfield
   {journal} {\bibinfo  {journal} {Marine and Freshwater Behaviour and
  Physiology}\ }\textbf {\bibinfo {volume} {41}},\ \bibinfo {pages} {91}
  (\bibinfo {year} {2008})}\BibitemShut {NoStop}%
\bibitem [{\citenamefont {Sidler}\ \emph {et~al.}(2018)\citenamefont {Sidler},
  \citenamefont {Michalec},\ and\ \citenamefont {Holzner}}]{sidler2018counter}%
  \BibitemOpen
  \bibfield  {author} {\bibinfo {author} {\bibfnamefont {D.}~\bibnamefont
  {Sidler}}, \bibinfo {author} {\bibfnamefont {F.}~\bibnamefont {Michalec}},\
  and\ \bibinfo {author} {\bibfnamefont {M.}~\bibnamefont {Holzner}},\
  }\bibfield  {title} {\bibinfo {title} {Counter-current swimming of lotic
  copepods as a possible mechanism for drift avoidance},\ }\href@noop {}
  {\bibfield  {journal} {\bibinfo  {journal} {Ecohydrology}\ }\textbf {\bibinfo
  {volume} {11}},\ \bibinfo {pages} {e1992} (\bibinfo {year}
  {2018})}\BibitemShut {NoStop}%
\bibitem [{\citenamefont {Elmi}\ \emph {et~al.}(2020)\citenamefont {Elmi},
  \citenamefont {Webster},\ and\ \citenamefont {Fields}}]{elmi2020the}%
  \BibitemOpen
  \bibfield  {author} {\bibinfo {author} {\bibfnamefont {D.}~\bibnamefont
  {Elmi}}, \bibinfo {author} {\bibfnamefont {D.~R.}\ \bibnamefont {Webster}},\
  and\ \bibinfo {author} {\bibfnamefont {D.~M.}\ \bibnamefont {Fields}},\
  }\bibfield  {title} {\bibinfo {title} {The response of the copepod acartia
  tonsa to the hydrodynamic cues of small-scale, dissipative eddies in
  turbulence},\ }\href@noop {} {\bibfield  {journal} {\bibinfo  {journal}
  {Journal of Experimental Biology}\ } (\bibinfo {year} {2020})}\BibitemShut
  {NoStop}%
\bibitem [{\citenamefont {Elmi}\ \emph {et~al.}(2022)\citenamefont {Elmi},
  \citenamefont {Webster},\ and\ \citenamefont {Fields}}]{elmi2022copepod}%
  \BibitemOpen
  \bibfield  {author} {\bibinfo {author} {\bibfnamefont {D.}~\bibnamefont
  {Elmi}}, \bibinfo {author} {\bibfnamefont {D.~R.}\ \bibnamefont {Webster}},\
  and\ \bibinfo {author} {\bibfnamefont {D.~M.}\ \bibnamefont {Fields}},\
  }\bibfield  {title} {\bibinfo {title} {Copepod interaction with small-scale,
  dissipative eddies in turbulence: Comparison among three marine species},\
  }\href@noop {} {\bibfield  {journal} {\bibinfo  {journal} {Limnology and
  Oceanography}\ }\textbf {\bibinfo {volume} {67}},\ \bibinfo {pages} {1820}
  (\bibinfo {year} {2022})}\BibitemShut {NoStop}%
\bibitem [{\citenamefont {Michalec}\ \emph {et~al.}(2015)\citenamefont
  {Michalec}, \citenamefont {Souissi},\ and\ \citenamefont
  {Holzner}}]{michalec2015turbulence}%
  \BibitemOpen
  \bibfield  {author} {\bibinfo {author} {\bibfnamefont {F.}~\bibnamefont
  {Michalec}}, \bibinfo {author} {\bibfnamefont {S.}~\bibnamefont {Souissi}},\
  and\ \bibinfo {author} {\bibfnamefont {M.}~\bibnamefont {Holzner}},\
  }\bibfield  {title} {\bibinfo {title} {Turbulence triggers vigorous swimming
  but hinders motion strategy in planktonic copepods},\ }\href@noop {}
  {\bibfield  {journal} {\bibinfo  {journal} {Journal of The Royal Society
  Interface}\ }\textbf {\bibinfo {volume} {12}},\ \bibinfo {pages} {20150158}
  (\bibinfo {year} {2015})}\BibitemShut {NoStop}%
\bibitem [{\citenamefont {Michalec}\ \emph {et~al.}(2017)\citenamefont
  {Michalec}, \citenamefont {Fouxon}, \citenamefont {Souissi},\ and\
  \citenamefont {Holzner}}]{michalec2017zooplankton}%
  \BibitemOpen
  \bibfield  {author} {\bibinfo {author} {\bibfnamefont {F.}~\bibnamefont
  {Michalec}}, \bibinfo {author} {\bibfnamefont {I.}~\bibnamefont {Fouxon}},
  \bibinfo {author} {\bibfnamefont {S.}~\bibnamefont {Souissi}},\ and\ \bibinfo
  {author} {\bibfnamefont {M.}~\bibnamefont {Holzner}},\ }\bibfield  {title}
  {\bibinfo {title} {Zooplankton can actively adjust their motility to
  turbulent flow},\ }\href@noop {} {\bibfield  {journal} {\bibinfo  {journal}
  {Proceedings of the National Academy of Sciences}\ }\textbf {\bibinfo
  {volume} {114}},\ \bibinfo {pages} {E11199} (\bibinfo {year}
  {2017})}\BibitemShut {NoStop}%
\bibitem [{\citenamefont {Adhikari}\ \emph {et~al.}(2015)\citenamefont
  {Adhikari}, \citenamefont {Gemmell}, \citenamefont {Hallberg}, \citenamefont
  {Longmire},\ and\ \citenamefont {Buskey}}]{adhikari2015simultaneous}%
  \BibitemOpen
  \bibfield  {author} {\bibinfo {author} {\bibfnamefont {D.}~\bibnamefont
  {Adhikari}}, \bibinfo {author} {\bibfnamefont {B.~J.}\ \bibnamefont
  {Gemmell}}, \bibinfo {author} {\bibfnamefont {M.~P.}\ \bibnamefont
  {Hallberg}}, \bibinfo {author} {\bibfnamefont {E.~K.}\ \bibnamefont
  {Longmire}},\ and\ \bibinfo {author} {\bibfnamefont {E.~J.}\ \bibnamefont
  {Buskey}},\ }\bibfield  {title} {\bibinfo {title} {Simultaneous measurement
  of 3d zooplankton trajectories and surrounding fluid velocity field in
  complex flows},\ }\href@noop {} {\bibfield  {journal} {\bibinfo  {journal}
  {Journal of Experimental Biology}\ } (\bibinfo {year} {2015})}\BibitemShut
  {NoStop}%
\bibitem [{\citenamefont {Colabrese}\ \emph {et~al.}(2017)\citenamefont
  {Colabrese}, \citenamefont {Gustavsson}, \citenamefont {Celani},\ and\
  \citenamefont {Biferale}}]{colabrese2017flow}%
  \BibitemOpen
  \bibfield  {author} {\bibinfo {author} {\bibfnamefont {S.}~\bibnamefont
  {Colabrese}}, \bibinfo {author} {\bibfnamefont {K.}~\bibnamefont
  {Gustavsson}}, \bibinfo {author} {\bibfnamefont {A.}~\bibnamefont {Celani}},\
  and\ \bibinfo {author} {\bibfnamefont {L.}~\bibnamefont {Biferale}},\
  }\bibfield  {title} {\bibinfo {title} {Flow navigation by smart microswimmers
  via reinforcement learning},\ }\href@noop {} {\bibfield  {journal} {\bibinfo
  {journal} {Physical Review Letters}\ }\textbf {\bibinfo {volume} {118}},\
  \bibinfo {pages} {158004} (\bibinfo {year} {2017})}\BibitemShut {NoStop}%
\bibitem [{\citenamefont {Biferale}\ \emph {et~al.}(2019)\citenamefont
  {Biferale}, \citenamefont {Bonaccorso}, \citenamefont {Buzzicotti},
  \citenamefont {Clark Di~Leoni},\ and\ \citenamefont
  {Gustavsson}}]{biferale2019zermelo}%
  \BibitemOpen
  \bibfield  {author} {\bibinfo {author} {\bibfnamefont {L.}~\bibnamefont
  {Biferale}}, \bibinfo {author} {\bibfnamefont {F.}~\bibnamefont
  {Bonaccorso}}, \bibinfo {author} {\bibfnamefont {M.}~\bibnamefont
  {Buzzicotti}}, \bibinfo {author} {\bibfnamefont {P.}~\bibnamefont {Clark
  Di~Leoni}},\ and\ \bibinfo {author} {\bibfnamefont {K.}~\bibnamefont
  {Gustavsson}},\ }\bibfield  {title} {\bibinfo {title} {Zermelo's problem:
  Optimal point-to-point navigation in 2d turbulent flows using reinforcement
  learning},\ }\href@noop {} {\bibfield  {journal} {\bibinfo  {journal} {Chaos:
  An Interdisciplinary Journal of Nonlinear Science}\ }\textbf {\bibinfo
  {volume} {29}},\ \bibinfo {pages} {103138} (\bibinfo {year}
  {2019})}\BibitemShut {NoStop}%
\bibitem [{\citenamefont {Schneider}\ and\ \citenamefont
  {Stark}(2019)}]{schneider2019optimal}%
  \BibitemOpen
  \bibfield  {author} {\bibinfo {author} {\bibfnamefont {E.}~\bibnamefont
  {Schneider}}\ and\ \bibinfo {author} {\bibfnamefont {H.}~\bibnamefont
  {Stark}},\ }\bibfield  {title} {\bibinfo {title} {Optimal steering of a smart
  active particle},\ }\href@noop {} {\bibfield  {journal} {\bibinfo  {journal}
  {EPL (Europhysics Letters)}\ }\textbf {\bibinfo {volume} {127}},\ \bibinfo
  {pages} {64003} (\bibinfo {year} {2019})}\BibitemShut {NoStop}%
\bibitem [{\citenamefont {Alageshan}\ \emph {et~al.}(2020)\citenamefont
  {Alageshan}, \citenamefont {Verma}, \citenamefont {Bec},\ and\ \citenamefont
  {Pandit}}]{alageshan2020machine}%
  \BibitemOpen
  \bibfield  {author} {\bibinfo {author} {\bibfnamefont {J.~K.}\ \bibnamefont
  {Alageshan}}, \bibinfo {author} {\bibfnamefont {A.~K.}\ \bibnamefont
  {Verma}}, \bibinfo {author} {\bibfnamefont {J.}~\bibnamefont {Bec}},\ and\
  \bibinfo {author} {\bibfnamefont {R.}~\bibnamefont {Pandit}},\ }\bibfield
  {title} {\bibinfo {title} {Machine learning strategies for path-planning
  microswimmers in turbulent flows},\ }\href@noop {} {\bibfield  {journal}
  {\bibinfo  {journal} {Physical Review E}\ }\textbf {\bibinfo {volume}
  {101}},\ \bibinfo {pages} {043110} (\bibinfo {year} {2020})}\BibitemShut
  {NoStop}%
\bibitem [{\citenamefont {Gunnarson}\ \emph {et~al.}(2021)\citenamefont
  {Gunnarson}, \citenamefont {Mandralis}, \citenamefont {Novati}, \citenamefont
  {Koumoutsakos},\ and\ \citenamefont {Dabiri}}]{gunnarson2021learning}%
  \BibitemOpen
  \bibfield  {author} {\bibinfo {author} {\bibfnamefont {P.}~\bibnamefont
  {Gunnarson}}, \bibinfo {author} {\bibfnamefont {I.}~\bibnamefont
  {Mandralis}}, \bibinfo {author} {\bibfnamefont {G.}~\bibnamefont {Novati}},
  \bibinfo {author} {\bibfnamefont {P.}~\bibnamefont {Koumoutsakos}},\ and\
  \bibinfo {author} {\bibfnamefont {J.~O.}\ \bibnamefont {Dabiri}},\ }\bibfield
   {title} {\bibinfo {title} {Learning efficient navigation in vortical flow
  fields},\ }\href@noop {} {\bibfield  {journal} {\bibinfo  {journal} {Nature
  Communications}\ }\textbf {\bibinfo {volume} {12}},\ \bibinfo {pages} {1}
  (\bibinfo {year} {2021})}\BibitemShut {NoStop}%
\bibitem [{\citenamefont {Qiu}\ \emph {et~al.}(2022{\natexlab{a}})\citenamefont
  {Qiu}, \citenamefont {Mousavi}, \citenamefont {Zhao},\ and\ \citenamefont
  {Gustavsson}}]{qiu2022active}%
  \BibitemOpen
  \bibfield  {author} {\bibinfo {author} {\bibfnamefont {J.}~\bibnamefont
  {Qiu}}, \bibinfo {author} {\bibfnamefont {N.}~\bibnamefont {Mousavi}},
  \bibinfo {author} {\bibfnamefont {L.}~\bibnamefont {Zhao}},\ and\ \bibinfo
  {author} {\bibfnamefont {K.}~\bibnamefont {Gustavsson}},\ }\bibfield  {title}
  {\bibinfo {title} {Active gyrotactic stability of microswimmers using
  hydromechanical signals},\ }\href@noop {} {\bibfield  {journal} {\bibinfo
  {journal} {Physical Review Fluids}\ }\textbf {\bibinfo {volume} {7}},\
  \bibinfo {pages} {014311} (\bibinfo {year} {2022}{\natexlab{a}})}\BibitemShut
  {NoStop}%
\bibitem [{\citenamefont {Xu}\ \emph {et~al.}(2023)\citenamefont {Xu},
  \citenamefont {Wu},\ and\ \citenamefont {Xi}}]{xu2023long}%
  \BibitemOpen
  \bibfield  {author} {\bibinfo {author} {\bibfnamefont {A.}~\bibnamefont
  {Xu}}, \bibinfo {author} {\bibfnamefont {H.}~\bibnamefont {Wu}},\ and\
  \bibinfo {author} {\bibfnamefont {H.}~\bibnamefont {Xi}},\ }\bibfield
  {title} {\bibinfo {title} {Long-distance migration with minimal energy
  consumption in a thermal turbulent environment},\ }\href@noop {} {\bibfield
  {journal} {\bibinfo  {journal} {Physical Review Fluids}\ }\textbf {\bibinfo
  {volume} {8}},\ \bibinfo {pages} {023502} (\bibinfo {year}
  {2023})}\BibitemShut {NoStop}%
\bibitem [{\citenamefont {Liebchen}\ and\ \citenamefont
  {Löwen}(2019)}]{liebchen2019optimal}%
  \BibitemOpen
  \bibfield  {author} {\bibinfo {author} {\bibfnamefont {B.}~\bibnamefont
  {Liebchen}}\ and\ \bibinfo {author} {\bibfnamefont {H.}~\bibnamefont
  {Löwen}},\ }\bibfield  {title} {\bibinfo {title} {Optimal navigation
  strategies for active particles},\ }\href@noop {} {\bibfield  {journal}
  {\bibinfo  {journal} {{EPL} (Europhysics Letters)}\ }\textbf {\bibinfo
  {volume} {127}},\ \bibinfo {pages} {34003} (\bibinfo {year}
  {2019})}\BibitemShut {NoStop}%
\bibitem [{\citenamefont {Daddi-Moussa-Ider}\ \emph {et~al.}(2021)\citenamefont
  {Daddi-Moussa-Ider}, \citenamefont {Löwen},\ and\ \citenamefont
  {Liebchen}}]{moussaIder2021hydrodynamics}%
  \BibitemOpen
  \bibfield  {author} {\bibinfo {author} {\bibfnamefont {A.}~\bibnamefont
  {Daddi-Moussa-Ider}}, \bibinfo {author} {\bibfnamefont {H.}~\bibnamefont
  {Löwen}},\ and\ \bibinfo {author} {\bibfnamefont {B.}~\bibnamefont
  {Liebchen}},\ }\bibfield  {title} {\bibinfo {title} {Hydrodynamics can
  determine the optimal route for microswimmer navigation},\ }\href@noop {}
  {\bibfield  {journal} {\bibinfo  {journal} {Communications Physics}\ }\textbf
  {\bibinfo {volume} {4}} (\bibinfo {year} {2021})}\BibitemShut {NoStop}%
\bibitem [{\citenamefont {Monthiller}\ \emph {et~al.}(2022)\citenamefont
  {Monthiller}, \citenamefont {Loisy}, \citenamefont {Koehl}, \citenamefont
  {Favier},\ and\ \citenamefont {Eloy}}]{monthiller2022surfing}%
  \BibitemOpen
  \bibfield  {author} {\bibinfo {author} {\bibfnamefont {R.}~\bibnamefont
  {Monthiller}}, \bibinfo {author} {\bibfnamefont {A.}~\bibnamefont {Loisy}},
  \bibinfo {author} {\bibfnamefont {M.~A.~R.}\ \bibnamefont {Koehl}}, \bibinfo
  {author} {\bibfnamefont {B.}~\bibnamefont {Favier}},\ and\ \bibinfo {author}
  {\bibfnamefont {C.}~\bibnamefont {Eloy}},\ }\bibfield  {title} {\bibinfo
  {title} {Surfing on turbulence: A strategy for planktonic navigation},\
  }\href@noop {} {\bibfield  {journal} {\bibinfo  {journal} {Physical Review
  Letters}\ }\textbf {\bibinfo {volume} {129}},\ \bibinfo {pages} {064502}
  (\bibinfo {year} {2022})}\BibitemShut {NoStop}%
\bibitem [{\citenamefont {Piro}\ \emph {et~al.}(2024)\citenamefont {Piro},
  \citenamefont {Vilfan}, \citenamefont {Golestanian},\ and\ \citenamefont
  {Mahault}}]{piro2023energetic}%
  \BibitemOpen
  \bibfield  {author} {\bibinfo {author} {\bibfnamefont {L.}~\bibnamefont
  {Piro}}, \bibinfo {author} {\bibfnamefont {A.}~\bibnamefont {Vilfan}},
  \bibinfo {author} {\bibfnamefont {R.}~\bibnamefont {Golestanian}},\ and\
  \bibinfo {author} {\bibfnamefont {B.}~\bibnamefont {Mahault}},\ }\bibfield
  {title} {\bibinfo {title} {Energetic cost of microswimmer navigation: The
  role of body shape},\ }\href@noop {} {\bibfield  {journal} {\bibinfo
  {journal} {Physical Review Research}\ }\textbf {\bibinfo {volume} {6}},\
  \bibinfo {pages} {013274} (\bibinfo {year} {2024})}\BibitemShut {NoStop}%
\bibitem [{\citenamefont {Mousavi}\ \emph {et~al.}(2024)\citenamefont
  {Mousavi}, \citenamefont {Qiu}, \citenamefont {Mehlig}, \citenamefont
  {Zhao},\ and\ \citenamefont {Gustavsson}}]{mousavi2024efficient}%
  \BibitemOpen
  \bibfield  {author} {\bibinfo {author} {\bibfnamefont {N.}~\bibnamefont
  {Mousavi}}, \bibinfo {author} {\bibfnamefont {J.}~\bibnamefont {Qiu}},
  \bibinfo {author} {\bibfnamefont {B.}~\bibnamefont {Mehlig}}, \bibinfo
  {author} {\bibfnamefont {L.}~\bibnamefont {Zhao}},\ and\ \bibinfo {author}
  {\bibfnamefont {K.}~\bibnamefont {Gustavsson}},\ }\bibfield  {title}
  {\bibinfo {title} {Efficient survival strategy for zooplankton in
  turbulence},\ }\href@noop {} {\bibfield  {journal} {\bibinfo  {journal}
  {Phys. Rev. Res.}\ }\textbf {\bibinfo {volume} {6}},\ \bibinfo {pages}
  {L022034} (\bibinfo {year} {2024})}\BibitemShut {NoStop}%
\bibitem [{\citenamefont {Qiu}\ \emph {et~al.}(2022{\natexlab{b}})\citenamefont
  {Qiu}, \citenamefont {Mousavi}, \citenamefont {Gustavsson}, \citenamefont
  {Xu}, \citenamefont {Mehlig},\ and\ \citenamefont
  {Zhao}}]{qiu2022navigation}%
  \BibitemOpen
  \bibfield  {author} {\bibinfo {author} {\bibfnamefont {J.}~\bibnamefont
  {Qiu}}, \bibinfo {author} {\bibfnamefont {N.}~\bibnamefont {Mousavi}},
  \bibinfo {author} {\bibfnamefont {K.}~\bibnamefont {Gustavsson}}, \bibinfo
  {author} {\bibfnamefont {C.}~\bibnamefont {Xu}}, \bibinfo {author}
  {\bibfnamefont {B.}~\bibnamefont {Mehlig}},\ and\ \bibinfo {author}
  {\bibfnamefont {L.}~\bibnamefont {Zhao}},\ }\bibfield  {title} {\bibinfo
  {title} {Navigation of micro-swimmers in steady flow: The importance of
  symmetries},\ }\href@noop {} {\bibfield  {journal} {\bibinfo  {journal}
  {Journal of Fluid Mechanics}\ }\textbf {\bibinfo {volume} {932}} (\bibinfo
  {year} {2022}{\natexlab{b}})}\BibitemShut {NoStop}%
\bibitem [{\citenamefont {Gustavsson}\ \emph {et~al.}(2017)\citenamefont
  {Gustavsson}, \citenamefont {Biferale}, \citenamefont {Celani},\ and\
  \citenamefont {Colabrese}}]{gustavsson2017finding}%
  \BibitemOpen
  \bibfield  {author} {\bibinfo {author} {\bibfnamefont {K.}~\bibnamefont
  {Gustavsson}}, \bibinfo {author} {\bibfnamefont {L.}~\bibnamefont
  {Biferale}}, \bibinfo {author} {\bibfnamefont {A.}~\bibnamefont {Celani}},\
  and\ \bibinfo {author} {\bibfnamefont {S.}~\bibnamefont {Colabrese}},\
  }\bibfield  {title} {\bibinfo {title} {Finding efficient swimming strategies
  in a three-dimensional chaotic flow by reinforcement learning},\ }\href@noop
  {} {\bibfield  {journal} {\bibinfo  {journal} {European Physical Journal E}\
  }\textbf {\bibinfo {volume} {40}},\ \bibinfo {pages} {110} (\bibinfo {year}
  {2017})}\BibitemShut {NoStop}%
\bibitem [{\citenamefont {Jakobsen}(2001)}]{jakobsen2001escape}%
  \BibitemOpen
  \bibfield  {author} {\bibinfo {author} {\bibfnamefont {H.~H.}\ \bibnamefont
  {Jakobsen}},\ }\bibfield  {title} {\bibinfo {title} {Escape response of
  planktonic protists to fluid mechanical signals},\ }\href@noop {} {\bibfield
  {journal} {\bibinfo  {journal} {Marine Ecology Progress Series}\ }\textbf
  {\bibinfo {volume} {214}},\ \bibinfo {pages} {67} (\bibinfo {year}
  {2001})}\BibitemShut {NoStop}%
\bibitem [{\citenamefont {Buskey}\ \emph {et~al.}(2002)\citenamefont {Buskey},
  \citenamefont {Lenz},\ and\ \citenamefont {Hartline}}]{buskey2002escape}%
  \BibitemOpen
  \bibfield  {author} {\bibinfo {author} {\bibfnamefont {E.}~\bibnamefont
  {Buskey}}, \bibinfo {author} {\bibfnamefont {P.}~\bibnamefont {Lenz}},\ and\
  \bibinfo {author} {\bibfnamefont {D.}~\bibnamefont {Hartline}},\ }\bibfield
  {title} {\bibinfo {title} {Escape behavior of planktonic copepods in response
  to hydrodynamic disturbances: high speed video analysis},\ }\href@noop {}
  {\bibfield  {journal} {\bibinfo  {journal} {Marine Ecology Progress Series}\
  }\textbf {\bibinfo {volume} {235}},\ \bibinfo {pages} {135} (\bibinfo {year}
  {2002})}\BibitemShut {NoStop}%
\bibitem [{\citenamefont {Ardeshiri}\ \emph {et~al.}(2016)\citenamefont
  {Ardeshiri}, \citenamefont {Benkeddad}, \citenamefont {Schmitt},
  \citenamefont {Souissi}, \citenamefont {Toschi},\ and\ \citenamefont
  {Calzavarini}}]{ardeshiri2016lagrangian}%
  \BibitemOpen
  \bibfield  {author} {\bibinfo {author} {\bibfnamefont {H.}~\bibnamefont
  {Ardeshiri}}, \bibinfo {author} {\bibfnamefont {I.}~\bibnamefont
  {Benkeddad}}, \bibinfo {author} {\bibfnamefont {F.~G.}\ \bibnamefont
  {Schmitt}}, \bibinfo {author} {\bibfnamefont {S.}~\bibnamefont {Souissi}},
  \bibinfo {author} {\bibfnamefont {F.}~\bibnamefont {Toschi}},\ and\ \bibinfo
  {author} {\bibfnamefont {E.}~\bibnamefont {Calzavarini}},\ }\bibfield
  {title} {\bibinfo {title} {Lagrangian model of copepod dynamics: Clustering
  by escape jumps in turbulence},\ }\href@noop {} {\bibfield  {journal}
  {\bibinfo  {journal} {Physical Review E}\ }\textbf {\bibinfo {volume} {93}},\
  \bibinfo {pages} {043117} (\bibinfo {year} {2016})}\BibitemShut {NoStop}%
\bibitem [{\citenamefont {Ardeshiri}\ \emph {et~al.}(2017)\citenamefont
  {Ardeshiri}, \citenamefont {Schmitt}, \citenamefont {Souissi}, \citenamefont
  {Toschi},\ and\ \citenamefont {Calzavarini}}]{ardeshiri2017copepods}%
  \BibitemOpen
  \bibfield  {author} {\bibinfo {author} {\bibfnamefont {H.}~\bibnamefont
  {Ardeshiri}}, \bibinfo {author} {\bibfnamefont {F.~G.}\ \bibnamefont
  {Schmitt}}, \bibinfo {author} {\bibfnamefont {S.}~\bibnamefont {Souissi}},
  \bibinfo {author} {\bibfnamefont {F.}~\bibnamefont {Toschi}},\ and\ \bibinfo
  {author} {\bibfnamefont {E.}~\bibnamefont {Calzavarini}},\ }\bibfield
  {title} {\bibinfo {title} {Copepods encounter rates from a model of escape
  jump behaviour in turbulence},\ }\href@noop {} {\bibfield  {journal}
  {\bibinfo  {journal} {Journal of Plankton Research}\ }\textbf {\bibinfo
  {volume} {39}},\ \bibinfo {pages} {878} (\bibinfo {year} {2017})}\BibitemShut
  {NoStop}%
\bibitem [{\citenamefont {Ki{\o}rboe}\ and\ \citenamefont
  {Visser}(1999)}]{kioerboe1999predator}%
  \BibitemOpen
  \bibfield  {author} {\bibinfo {author} {\bibfnamefont {T.}~\bibnamefont
  {Ki{\o}rboe}}\ and\ \bibinfo {author} {\bibfnamefont {A.~W.}\ \bibnamefont
  {Visser}},\ }\bibfield  {title} {\bibinfo {title} {Predator and prey
  perception in copepods due to hydromechanical signals},\ }\href@noop {}
  {\bibfield  {journal} {\bibinfo  {journal} {Marine Ecology Progress Series}\
  }\textbf {\bibinfo {volume} {179}},\ \bibinfo {pages} {81} (\bibinfo {year}
  {1999})}\BibitemShut {NoStop}%
\bibitem [{\citenamefont {Recht}(2019)}]{recht2019tour}%
  \BibitemOpen
  \bibfield  {author} {\bibinfo {author} {\bibfnamefont {B.}~\bibnamefont
  {Recht}},\ }\bibfield  {title} {\bibinfo {title} {A tour of reinforcement
  learning: The view from continuous control},\ }\href@noop {} {\bibfield
  {journal} {\bibinfo  {journal} {Annual Review of Control, Robotics, and
  Autonomous Systems}\ }\textbf {\bibinfo {volume} {2}},\ \bibinfo {pages}
  {253} (\bibinfo {year} {2019})}\BibitemShut {NoStop}%
\bibitem [{\citenamefont {Cichos}\ \emph {et~al.}(2020)\citenamefont {Cichos},
  \citenamefont {Gustavsson}, \citenamefont {Mehlig},\ and\ \citenamefont
  {Volpe}}]{cichos2020machine}%
  \BibitemOpen
  \bibfield  {author} {\bibinfo {author} {\bibfnamefont {F.}~\bibnamefont
  {Cichos}}, \bibinfo {author} {\bibfnamefont {K.}~\bibnamefont {Gustavsson}},
  \bibinfo {author} {\bibfnamefont {B.}~\bibnamefont {Mehlig}},\ and\ \bibinfo
  {author} {\bibfnamefont {G.}~\bibnamefont {Volpe}},\ }\bibfield  {title}
  {\bibinfo {title} {Machine learning for active matter},\ }\href@noop {}
  {\bibfield  {journal} {\bibinfo  {journal} {Nature Machine Intelligence}\
  }\textbf {\bibinfo {volume} {2}},\ \bibinfo {pages} {94} (\bibinfo {year}
  {2020})}\BibitemShut {NoStop}%
\bibitem [{\citenamefont {Tsang}\ \emph {et~al.}(2020)\citenamefont {Tsang},
  \citenamefont {Demir}, \citenamefont {Ding},\ and\ \citenamefont
  {Pak}}]{tsang2020roads}%
  \BibitemOpen
  \bibfield  {author} {\bibinfo {author} {\bibfnamefont {A.~C.~H.}\
  \bibnamefont {Tsang}}, \bibinfo {author} {\bibfnamefont {E.}~\bibnamefont
  {Demir}}, \bibinfo {author} {\bibfnamefont {Y.}~\bibnamefont {Ding}},\ and\
  \bibinfo {author} {\bibfnamefont {O.~S.}\ \bibnamefont {Pak}},\ }\bibfield
  {title} {\bibinfo {title} {Roads to smart artificial microswimmers},\
  }\href@noop {} {\bibfield  {journal} {\bibinfo  {journal} {Advanced
  Intelligent Systems}\ }\textbf {\bibinfo {volume} {2}} (\bibinfo {year}
  {2020})}\BibitemShut {NoStop}%
\bibitem [{\citenamefont {Mui{\~n}os-Landin}\ \emph {et~al.}(2021)\citenamefont
  {Mui{\~n}os-Landin}, \citenamefont {Fischer}, \citenamefont {Holubec},\ and\
  \citenamefont {Cichos}}]{muinos2021reinforcement}%
  \BibitemOpen
  \bibfield  {author} {\bibinfo {author} {\bibfnamefont {S.}~\bibnamefont
  {Mui{\~n}os-Landin}}, \bibinfo {author} {\bibfnamefont {A.}~\bibnamefont
  {Fischer}}, \bibinfo {author} {\bibfnamefont {V.}~\bibnamefont {Holubec}},\
  and\ \bibinfo {author} {\bibfnamefont {F.}~\bibnamefont {Cichos}},\
  }\bibfield  {title} {\bibinfo {title} {Reinforcement learning with artificial
  microswimmers},\ }\href@noop {} {\bibfield  {journal} {\bibinfo  {journal}
  {Science Robotics}\ }\textbf {\bibinfo {volume} {6}},\ \bibinfo {pages}
  {eabd9285} (\bibinfo {year} {2021})}\BibitemShut {NoStop}%
\bibitem [{\citenamefont {Zou}\ \emph {et~al.}(2022)\citenamefont {Zou},
  \citenamefont {Liu}, \citenamefont {Young}, \citenamefont {Pak},\ and\
  \citenamefont {Tsang}}]{zou2022gait}%
  \BibitemOpen
  \bibfield  {author} {\bibinfo {author} {\bibfnamefont {Z.}~\bibnamefont
  {Zou}}, \bibinfo {author} {\bibfnamefont {Y.}~\bibnamefont {Liu}}, \bibinfo
  {author} {\bibfnamefont {Y.-N.}\ \bibnamefont {Young}}, \bibinfo {author}
  {\bibfnamefont {O.~S.}\ \bibnamefont {Pak}},\ and\ \bibinfo {author}
  {\bibfnamefont {A.~C.~H.}\ \bibnamefont {Tsang}},\ }\bibfield  {title}
  {\bibinfo {title} {Gait switching and targeted navigation of microswimmers
  via deep reinforcement learning},\ }\href@noop {} {\bibfield  {journal}
  {\bibinfo  {journal} {Communications Physics}\ }\textbf {\bibinfo {volume}
  {5}} (\bibinfo {year} {2022})}\BibitemShut {NoStop}%
\bibitem [{\citenamefont {Mo}\ \emph {et~al.}(2023)\citenamefont {Mo},
  \citenamefont {Li},\ and\ \citenamefont {Bian}}]{mo2023challenges}%
  \BibitemOpen
  \bibfield  {author} {\bibinfo {author} {\bibfnamefont {C.}~\bibnamefont
  {Mo}}, \bibinfo {author} {\bibfnamefont {G.}~\bibnamefont {Li}},\ and\
  \bibinfo {author} {\bibfnamefont {X.}~\bibnamefont {Bian}},\ }\bibfield
  {title} {\bibinfo {title} {Challenges and attempts to make intelligent
  microswimmers},\ }\href@noop {} {\bibfield  {journal} {\bibinfo  {journal}
  {Frontiers in Physics}\ }\textbf {\bibinfo {volume} {11}},\ \bibinfo {pages}
  {1279883} (\bibinfo {year} {2023})}\BibitemShut {NoStop}%
\bibitem [{\citenamefont {Rey}\ \emph {et~al.}(2023)\citenamefont {Rey},
  \citenamefont {Volpe},\ and\ \citenamefont {Volpe}}]{rey2023light}%
  \BibitemOpen
  \bibfield  {author} {\bibinfo {author} {\bibfnamefont {M.}~\bibnamefont
  {Rey}}, \bibinfo {author} {\bibfnamefont {G.}~\bibnamefont {Volpe}},\ and\
  \bibinfo {author} {\bibfnamefont {G.}~\bibnamefont {Volpe}},\ }\bibfield
  {title} {\bibinfo {title} {Light, matter, action: Shining light on active
  matter},\ }\href@noop {} {\bibfield  {journal} {\bibinfo  {journal} {ACS
  photonics}\ }\textbf {\bibinfo {volume} {10}},\ \bibinfo {pages} {1188}
  (\bibinfo {year} {2023})}\BibitemShut {NoStop}%
\bibitem [{\citenamefont {Amoudruz}\ \emph {et~al.}(2024)\citenamefont
  {Amoudruz}, \citenamefont {Litvinov},\ and\ \citenamefont
  {Koumoutsakos}}]{amoudruz2024path}%
  \BibitemOpen
  \bibfield  {author} {\bibinfo {author} {\bibfnamefont {L.}~\bibnamefont
  {Amoudruz}}, \bibinfo {author} {\bibfnamefont {S.}~\bibnamefont {Litvinov}},\
  and\ \bibinfo {author} {\bibfnamefont {P.}~\bibnamefont {Koumoutsakos}},\
  }\bibfield  {title} {\bibinfo {title} {Path planning of magnetic
  microswimmers in high-fidelity simulations of capillaries with deep
  reinforcement learning},\ }\href@noop {} {\bibfield  {journal} {\bibinfo
  {journal} {arXiv preprint arXiv:2404.02171}\ } (\bibinfo {year}
  {2024})}\BibitemShut {NoStop}%
\bibitem [{\citenamefont {Pradip}\ and\ \citenamefont
  {Cichos}(2022)}]{pradip2022deep}%
  \BibitemOpen
  \bibfield  {author} {\bibinfo {author} {\bibfnamefont {R.}~\bibnamefont
  {Pradip}}\ and\ \bibinfo {author} {\bibfnamefont {F.}~\bibnamefont
  {Cichos}},\ }\bibfield  {title} {\bibinfo {title} {Deep reinforcement
  learning with artificial microswimmers},\ }in\ \href@noop {} {\emph {\bibinfo
  {booktitle} {Emerging Topics in Artificial Intelligence (ETAI) 2022}}},\
  Vol.\ \bibinfo {volume} {12204}\ (\bibinfo {organization} {SPIE},\ \bibinfo
  {year} {2022})\ pp.\ \bibinfo {pages} {104--110}\BibitemShut {NoStop}%
\bibitem [{\citenamefont {Schrage}\ \emph {et~al.}(2023)\citenamefont
  {Schrage}, \citenamefont {Medany},\ and\ \citenamefont
  {Ahmed}}]{schrage2023ultrasound}%
  \BibitemOpen
  \bibfield  {author} {\bibinfo {author} {\bibfnamefont {M.}~\bibnamefont
  {Schrage}}, \bibinfo {author} {\bibfnamefont {M.}~\bibnamefont {Medany}},\
  and\ \bibinfo {author} {\bibfnamefont {D.}~\bibnamefont {Ahmed}},\ }\bibfield
   {title} {\bibinfo {title} {Ultrasound microrobots with reinforcement
  learning},\ }\href@noop {} {\bibfield  {journal} {\bibinfo  {journal}
  {Advanced Materials Technologies}\ }\textbf {\bibinfo {volume} {8}},\
  \bibinfo {pages} {2201702} (\bibinfo {year} {2023})}\BibitemShut {NoStop}%
\bibitem [{\citenamefont {Durham}\ \emph {et~al.}(2009)\citenamefont {Durham},
  \citenamefont {Kessler},\ and\ \citenamefont
  {Stocker}}]{durham2009disruption}%
  \BibitemOpen
  \bibfield  {author} {\bibinfo {author} {\bibfnamefont {W.~M.}\ \bibnamefont
  {Durham}}, \bibinfo {author} {\bibfnamefont {J.~O.}\ \bibnamefont
  {Kessler}},\ and\ \bibinfo {author} {\bibfnamefont {R.}~\bibnamefont
  {Stocker}},\ }\bibfield  {title} {\bibinfo {title} {Disruption of vertical
  motility by shear triggers formation of thin phytoplankton layers},\
  }\href@noop {} {\bibfield  {journal} {\bibinfo  {journal} {Science}\ }\textbf
  {\bibinfo {volume} {323}},\ \bibinfo {pages} {1067} (\bibinfo {year}
  {2009})}\BibitemShut {NoStop}%
\bibitem [{\citenamefont {Li}\ \emph {et~al.}(2008)\citenamefont {Li},
  \citenamefont {Perlman}, \citenamefont {Wan}, \citenamefont {Yang},
  \citenamefont {Meneveau}, \citenamefont {Burns}, \citenamefont {Chen},
  \citenamefont {Szalay},\ and\ \citenamefont {Eyink}}]{JohnsHopkins}%
  \BibitemOpen
  \bibfield  {author} {\bibinfo {author} {\bibfnamefont {Y.}~\bibnamefont
  {Li}}, \bibinfo {author} {\bibfnamefont {E.}~\bibnamefont {Perlman}},
  \bibinfo {author} {\bibfnamefont {M.}~\bibnamefont {Wan}}, \bibinfo {author}
  {\bibfnamefont {Y.}~\bibnamefont {Yang}}, \bibinfo {author} {\bibfnamefont
  {C.}~\bibnamefont {Meneveau}}, \bibinfo {author} {\bibfnamefont
  {R.}~\bibnamefont {Burns}}, \bibinfo {author} {\bibfnamefont
  {S.}~\bibnamefont {Chen}}, \bibinfo {author} {\bibfnamefont {A.}~\bibnamefont
  {Szalay}},\ and\ \bibinfo {author} {\bibfnamefont {G.}~\bibnamefont
  {Eyink}},\ }\bibfield  {title} {\bibinfo {title} {A public turbulence
  database cluster and applications to study lagrangian evolution of velocity
  increments in turbulence},\ }\href
  {https://doi.org/10.1080/14685240802376389} {\bibfield  {journal} {\bibinfo
  {journal} {Journal of Turbulence}\ }\textbf {\bibinfo {volume} {9}},\
  \bibinfo {pages} {N31} (\bibinfo {year} {2008})}\BibitemShut {NoStop}%
\bibitem [{\citenamefont {Perlman}\ \emph {et~al.}(2007)\citenamefont
  {Perlman}, \citenamefont {Burns}, \citenamefont {Li},\ and\ \citenamefont
  {Meneveau}}]{JohnsHopkins2}%
  \BibitemOpen
  \bibfield  {author} {\bibinfo {author} {\bibfnamefont {E.}~\bibnamefont
  {Perlman}}, \bibinfo {author} {\bibfnamefont {R.}~\bibnamefont {Burns}},
  \bibinfo {author} {\bibfnamefont {Y.}~\bibnamefont {Li}},\ and\ \bibinfo
  {author} {\bibfnamefont {C.}~\bibnamefont {Meneveau}},\ }\bibfield  {title}
  {\bibinfo {title} {Data exploration of turbulence simulations using a
  database cluster},\ }in\ \href {https://doi.org/10.1145/1362622.1362654}
  {\emph {\bibinfo {booktitle} {Proceedings of the 2007 ACM/IEEE conference on
  Supercomputing}}},\ \bibinfo {series and number} {SC ’07}\ (\bibinfo
  {publisher} {ACM},\ \bibinfo {year} {2007})\BibitemShut {NoStop}%
\bibitem [{\citenamefont {Gustavsson}\ and\ \citenamefont
  {Mehlig}(2016)}]{gustavsson2016statistical}%
  \BibitemOpen
  \bibfield  {author} {\bibinfo {author} {\bibfnamefont {K.}~\bibnamefont
  {Gustavsson}}\ and\ \bibinfo {author} {\bibfnamefont {B.}~\bibnamefont
  {Mehlig}},\ }\bibfield  {title} {\bibinfo {title} {{Statistical models for
  spatial patterns of heavy particles in turbulence}},\ }\href@noop {}
  {\bibfield  {journal} {\bibinfo  {journal} {Advances in Physics}\ }\textbf
  {\bibinfo {volume} {65}},\ \bibinfo {pages} {1} (\bibinfo {year} {2016})},\
  \Eprint {https://arxiv.org/abs/1412.4374} {1412.4374} \BibitemShut {NoStop}%
\bibitem [{\citenamefont {Bec}\ \emph {et~al.}(2024)\citenamefont {Bec},
  \citenamefont {Gustavsson},\ and\ \citenamefont
  {Mehlig}}]{bec2024statistical}%
  \BibitemOpen
  \bibfield  {author} {\bibinfo {author} {\bibfnamefont {J.}~\bibnamefont
  {Bec}}, \bibinfo {author} {\bibfnamefont {K.}~\bibnamefont {Gustavsson}},\
  and\ \bibinfo {author} {\bibfnamefont {B.}~\bibnamefont {Mehlig}},\
  }\bibfield  {title} {\bibinfo {title} {Statistical models for the dynamics of
  heavy particles in turbulence},\ }\href@noop {} {\bibfield  {journal}
  {\bibinfo  {journal} {Annual Review of Fluid Mechanics}\ }\textbf {\bibinfo
  {volume} {56}},\ \bibinfo {pages} {1} (\bibinfo {year} {2024})}\BibitemShut
  {NoStop}%
\bibitem [{\citenamefont {Wilkinson}\ \emph {et~al.}(2007)\citenamefont
  {Wilkinson}, \citenamefont {Mehlig}, \citenamefont {{\"O}stlund},\ and\
  \citenamefont {Duncan}}]{wilkinson2007unmixing}%
  \BibitemOpen
  \bibfield  {author} {\bibinfo {author} {\bibfnamefont {M.}~\bibnamefont
  {Wilkinson}}, \bibinfo {author} {\bibfnamefont {B.}~\bibnamefont {Mehlig}},
  \bibinfo {author} {\bibfnamefont {S.}~\bibnamefont {{\"O}stlund}},\ and\
  \bibinfo {author} {\bibfnamefont {K.~P.}\ \bibnamefont {Duncan}},\ }\bibfield
   {title} {\bibinfo {title} {Unmixing in random flows},\ }\href@noop {}
  {\bibfield  {journal} {\bibinfo  {journal} {Physiscs of Fluids}\ }\textbf
  {\bibinfo {volume} {19}},\ \bibinfo {pages} {113303} (\bibinfo {year}
  {2007})}\BibitemShut {NoStop}%
\bibitem [{\citenamefont {Duncan}\ \emph {et~al.}(2005)\citenamefont {Duncan},
  \citenamefont {Mehlig}, \citenamefont {{\"O}stlund},\ and\ \citenamefont
  {Wilkinson}}]{duncan2005clustering}%
  \BibitemOpen
  \bibfield  {author} {\bibinfo {author} {\bibfnamefont {K.~P.}\ \bibnamefont
  {Duncan}}, \bibinfo {author} {\bibfnamefont {B.}~\bibnamefont {Mehlig}},
  \bibinfo {author} {\bibfnamefont {S.}~\bibnamefont {{\"O}stlund}},\ and\
  \bibinfo {author} {\bibfnamefont {M.}~\bibnamefont {Wilkinson}},\ }\bibfield
  {title} {\bibinfo {title} {Clustering by mixing flows},\ }\href@noop {}
  {\bibfield  {journal} {\bibinfo  {journal} {Physical Review Letters}\
  }\textbf {\bibinfo {volume} {95}},\ \bibinfo {pages} {240602} (\bibinfo
  {year} {2005})}\BibitemShut {NoStop}%
\bibitem [{\citenamefont {Borgnino}\ \emph {et~al.}(2019)\citenamefont
  {Borgnino}, \citenamefont {Gustavsson}, \citenamefont {De~Lillo},
  \citenamefont {Boffetta}, \citenamefont {Cencini},\ and\ \citenamefont
  {Mehlig}}]{borgnino2019alignment}%
  \BibitemOpen
  \bibfield  {author} {\bibinfo {author} {\bibfnamefont {M.}~\bibnamefont
  {Borgnino}}, \bibinfo {author} {\bibfnamefont {K.}~\bibnamefont
  {Gustavsson}}, \bibinfo {author} {\bibfnamefont {F.}~\bibnamefont
  {De~Lillo}}, \bibinfo {author} {\bibfnamefont {G.}~\bibnamefont {Boffetta}},
  \bibinfo {author} {\bibfnamefont {M.}~\bibnamefont {Cencini}},\ and\ \bibinfo
  {author} {\bibfnamefont {B.}~\bibnamefont {Mehlig}},\ }\bibfield  {title}
  {\bibinfo {title} {Alignment of nonspherical active particles in chaotic
  flows},\ }\href@noop {} {\bibfield  {journal} {\bibinfo  {journal} {Physical
  Review Letters}\ }\textbf {\bibinfo {volume} {123}},\ \bibinfo {pages}
  {138003} (\bibinfo {year} {2019})}\BibitemShut {NoStop}%
\bibitem [{\citenamefont {Svetlichny}\ \emph {et~al.}(2020)\citenamefont
  {Svetlichny}, \citenamefont {Larsen},\ and\ \citenamefont
  {Ki{\o}rboe}}]{svetlichny2020kinematic}%
  \BibitemOpen
  \bibfield  {author} {\bibinfo {author} {\bibfnamefont {L.}~\bibnamefont
  {Svetlichny}}, \bibinfo {author} {\bibfnamefont {P.~S.}\ \bibnamefont
  {Larsen}},\ and\ \bibinfo {author} {\bibfnamefont {T.}~\bibnamefont
  {Ki{\o}rboe}},\ }\bibfield  {title} {\bibinfo {title} {Kinematic and dynamic
  scaling of copepod swimming},\ }\href@noop {} {\bibfield  {journal} {\bibinfo
   {journal} {Fluids}\ }\textbf {\bibinfo {volume} {5}},\ \bibinfo {pages} {68}
  (\bibinfo {year} {2020})}\BibitemShut {NoStop}%
\bibitem [{\citenamefont {Yamazaki}\ and\ \citenamefont
  {Squires}(1996)}]{yamazaki1996comparison}%
  \BibitemOpen
  \bibfield  {author} {\bibinfo {author} {\bibfnamefont {H.}~\bibnamefont
  {Yamazaki}}\ and\ \bibinfo {author} {\bibfnamefont {K.~D.}\ \bibnamefont
  {Squires}},\ }\bibfield  {title} {\bibinfo {title} {Comparison of oceanic
  turbulence and copepod swimming},\ }\href@noop {} {\bibfield  {journal}
  {\bibinfo  {journal} {Marine Ecology Progress Series}\ }\textbf {\bibinfo
  {volume} {144}},\ \bibinfo {pages} {299} (\bibinfo {year}
  {1996})}\BibitemShut {NoStop}%
\bibitem [{\citenamefont {Fuchs}\ and\ \citenamefont
  {Gerbi}(2016)}]{fuchs2016seascape}%
  \BibitemOpen
  \bibfield  {author} {\bibinfo {author} {\bibfnamefont {H.~L.}\ \bibnamefont
  {Fuchs}}\ and\ \bibinfo {author} {\bibfnamefont {G.~P.}\ \bibnamefont
  {Gerbi}},\ }\bibfield  {title} {\bibinfo {title} {Seascape-level variation in
  turbulence-and wave-generated hydrodynamic signals experienced by plankton},\
  }\href@noop {} {\bibfield  {journal} {\bibinfo  {journal} {Progress in
  Oceanography}\ }\textbf {\bibinfo {volume} {141}},\ \bibinfo {pages} {109}
  (\bibinfo {year} {2016})}\BibitemShut {NoStop}%
\bibitem [{\citenamefont {Sutton}\ and\ \citenamefont
  {Barto}(2018)}]{sutton2018reinforcement}%
  \BibitemOpen
  \bibfield  {author} {\bibinfo {author} {\bibfnamefont {R.~S.}\ \bibnamefont
  {Sutton}}\ and\ \bibinfo {author} {\bibfnamefont {A.~G.}\ \bibnamefont
  {Barto}},\ }\href@noop {} {\emph {\bibinfo {title} {Reinforcement learning:
  An introduction}}}\ (\bibinfo  {publisher} {MIT press},\ \bibinfo {year}
  {2018})\BibitemShut {NoStop}%
\bibitem [{\citenamefont {Mehlig}(2021)}]{mehlig2021machine}%
  \BibitemOpen
  \bibfield  {author} {\bibinfo {author} {\bibfnamefont {B.}~\bibnamefont
  {Mehlig}},\ }\href@noop {} {\emph {\bibinfo {title} {Machine Learning with
  Neural Networks: An Introduction for Scientists and Engineers}}}\ (\bibinfo
  {publisher} {Cambridge University Press},\ \bibinfo {year}
  {2021})\BibitemShut {NoStop}%
\bibitem [{\citenamefont {Gilbert}\ and\ \citenamefont
  {Buskey}(2005)}]{gilbert2005turbulence}%
  \BibitemOpen
  \bibfield  {author} {\bibinfo {author} {\bibfnamefont {O.~M.}\ \bibnamefont
  {Gilbert}}\ and\ \bibinfo {author} {\bibfnamefont {E.~J.}\ \bibnamefont
  {Buskey}},\ }\bibfield  {title} {\bibinfo {title} {Turbulence decreases the
  hydrodynamic predator sensing ability of the calanoid copepod acartia
  tonsa},\ }\href@noop {} {\bibfield  {journal} {\bibinfo  {journal} {Journal
  of Plankton Research}\ }\textbf {\bibinfo {volume} {27}},\ \bibinfo {pages}
  {1067} (\bibinfo {year} {2005})}\BibitemShut {NoStop}%
\bibitem [{\citenamefont {Calascibetta}\ \emph {et~al.}(2023)\citenamefont
  {Calascibetta}, \citenamefont {Biferale}, \citenamefont {Borra},
  \citenamefont {Celani},\ and\ \citenamefont
  {Cencini}}]{calascibetta2023taming}%
  \BibitemOpen
  \bibfield  {author} {\bibinfo {author} {\bibfnamefont {C.}~\bibnamefont
  {Calascibetta}}, \bibinfo {author} {\bibfnamefont {L.}~\bibnamefont
  {Biferale}}, \bibinfo {author} {\bibfnamefont {F.}~\bibnamefont {Borra}},
  \bibinfo {author} {\bibfnamefont {A.}~\bibnamefont {Celani}},\ and\ \bibinfo
  {author} {\bibfnamefont {M.}~\bibnamefont {Cencini}},\ }\bibfield  {title}
  {\bibinfo {title} {Taming lagrangian chaos with multi-objective reinforcement
  learning},\ }\href@noop {} {\bibfield  {journal} {\bibinfo  {journal} {The
  European Physical Journal E}\ }\textbf {\bibinfo {volume} {46}} (\bibinfo
  {year} {2023})}\BibitemShut {NoStop}%
\bibitem [{\citenamefont {Machiels}(1997)}]{machiels1997predictability}%
  \BibitemOpen
  \bibfield  {author} {\bibinfo {author} {\bibfnamefont {L.}~\bibnamefont
  {Machiels}},\ }\bibfield  {title} {\bibinfo {title} {Predictability of
  small-scale motion in isotropic fluid turbulence},\ }\href@noop {} {\bibfield
   {journal} {\bibinfo  {journal} {Physical Review Letters}\ }\textbf {\bibinfo
  {volume} {79}},\ \bibinfo {pages} {3411} (\bibinfo {year}
  {1997})}\BibitemShut {NoStop}%
\bibitem [{\citenamefont {Borgnino}\ \emph {et~al.}(2022)\citenamefont
  {Borgnino}, \citenamefont {Boffetta}, \citenamefont {Cencini}, \citenamefont
  {De~Lillo},\ and\ \citenamefont {Gustavsson}}]{borgnino2022alignment}%
  \BibitemOpen
  \bibfield  {author} {\bibinfo {author} {\bibfnamefont {M.}~\bibnamefont
  {Borgnino}}, \bibinfo {author} {\bibfnamefont {G.}~\bibnamefont {Boffetta}},
  \bibinfo {author} {\bibfnamefont {M.}~\bibnamefont {Cencini}}, \bibinfo
  {author} {\bibfnamefont {F.}~\bibnamefont {De~Lillo}},\ and\ \bibinfo
  {author} {\bibfnamefont {K.}~\bibnamefont {Gustavsson}},\ }\bibfield  {title}
  {\bibinfo {title} {Alignment of elongated swimmers in a laminar and turbulent
  kolmogorov flow},\ }\href@noop {} {\bibfield  {journal} {\bibinfo  {journal}
  {Physical Review Fluids}\ }\textbf {\bibinfo {volume} {7}},\ \bibinfo {pages}
  {074603} (\bibinfo {year} {2022})}\BibitemShut {NoStop}%
\bibitem [{\citenamefont {J.~Meibohm}\ and\ \citenamefont
  {Gustavsson}(2024)}]{meibohm2024caustic}%
  \BibitemOpen
  \bibfield  {author} {\bibinfo {author} {\bibfnamefont {B.~M.}\ \bibnamefont
  {J.~Meibohm}, \bibfnamefont {L.~Sundberg}}\ and\ \bibinfo {author}
  {\bibfnamefont {K.}~\bibnamefont {Gustavsson}},\ }\bibfield  {title}
  {\bibinfo {title} {Caustic formation in a non-gaussian model for turbulent
  aerosols},\ }\href@noop {} {\bibfield  {journal} {\bibinfo  {journal}
  {Physical Review Fluids}\ }\textbf {\bibinfo {volume} {9}},\ \bibinfo {pages}
  {024302} (\bibinfo {year} {2024})}\BibitemShut {NoStop}%
\bibitem [{\citenamefont {Mnih}\ \emph {et~al.}(2015)\citenamefont {Mnih},
  \citenamefont {Kavukcuoglu}, \citenamefont {Silver}, \citenamefont {Rusu},
  \citenamefont {Veness}, \citenamefont {Bellemare}, \citenamefont {Graves},
  \citenamefont {Riedmiller}, \citenamefont {Fidjeland}, \citenamefont
  {Ostrovski} \emph {et~al.}}]{mnih2015human}%
  \BibitemOpen
  \bibfield  {author} {\bibinfo {author} {\bibfnamefont {V.}~\bibnamefont
  {Mnih}}, \bibinfo {author} {\bibfnamefont {K.}~\bibnamefont {Kavukcuoglu}},
  \bibinfo {author} {\bibfnamefont {D.}~\bibnamefont {Silver}}, \bibinfo
  {author} {\bibfnamefont {A.~A.}\ \bibnamefont {Rusu}}, \bibinfo {author}
  {\bibfnamefont {J.}~\bibnamefont {Veness}}, \bibinfo {author} {\bibfnamefont
  {M.~G.}\ \bibnamefont {Bellemare}}, \bibinfo {author} {\bibfnamefont
  {A.}~\bibnamefont {Graves}}, \bibinfo {author} {\bibfnamefont
  {M.}~\bibnamefont {Riedmiller}}, \bibinfo {author} {\bibfnamefont {A.~K.}\
  \bibnamefont {Fidjeland}}, \bibinfo {author} {\bibfnamefont {G.}~\bibnamefont
  {Ostrovski}}, \emph {et~al.},\ }\bibfield  {title} {\bibinfo {title}
  {Human-level control through deep reinforcement learning},\ }\href@noop {}
  {\bibfield  {journal} {\bibinfo  {journal} {nature}\ }\textbf {\bibinfo
  {volume} {518}},\ \bibinfo {pages} {529} (\bibinfo {year}
  {2015})}\BibitemShut {NoStop}%
\bibitem [{\citenamefont {Lillicrap}\ \emph {et~al.}(2015)\citenamefont
  {Lillicrap}, \citenamefont {Hunt}, \citenamefont {Pritzel}, \citenamefont
  {Heess}, \citenamefont {Erez}, \citenamefont {Tassa}, \citenamefont
  {Silver},\ and\ \citenamefont {Wierstra}}]{lillicrap2015continuous}%
  \BibitemOpen
  \bibfield  {author} {\bibinfo {author} {\bibfnamefont {T.~P.}\ \bibnamefont
  {Lillicrap}}, \bibinfo {author} {\bibfnamefont {J.~J.}\ \bibnamefont {Hunt}},
  \bibinfo {author} {\bibfnamefont {A.}~\bibnamefont {Pritzel}}, \bibinfo
  {author} {\bibfnamefont {N.}~\bibnamefont {Heess}}, \bibinfo {author}
  {\bibfnamefont {T.}~\bibnamefont {Erez}}, \bibinfo {author} {\bibfnamefont
  {Y.}~\bibnamefont {Tassa}}, \bibinfo {author} {\bibfnamefont
  {D.}~\bibnamefont {Silver}},\ and\ \bibinfo {author} {\bibfnamefont
  {D.}~\bibnamefont {Wierstra}},\ }\bibfield  {title} {\bibinfo {title}
  {Continuous control with deep reinforcement learning},\ }\href@noop {} {\
  (\bibinfo {year} {2015})},\ \Eprint {https://arxiv.org/abs/1509.02971}
  {arXiv:1509.02971 [cs.LG]} \BibitemShut {NoStop}%
\end{thebibliography}


\end{document}